\documentclass[a4paper,12pt]{article}
\usepackage{amsmath}
\usepackage{amssymb}
\usepackage{fullpage}
\usepackage{verbatim}

\usepackage[T1,OT1]{fontenc}

\usepackage{graphicx}

\allowdisplaybreaks[3]

\usepackage[hang]{footmisc}
\usepackage[compress,square,numbers]{natbib}
\usepackage[colorlinks,linktocpage,linkcolor=blue,citecolor=blue,urlcolor=blue]{hyperref}

\renewcommand{\d}{\mathrm{d}}

\newcommand{\w}{\wedge}

\begin{document}

\numberwithin{equation}{section}

\thispagestyle{empty}

\vspace*{1cm}

\begin{center}

{\LARGE \bf Control issues of KKLT}

\vspace{1.5cm}
{\large Xin Gao$^{1,2}$, Arthur Hebecker$^1$ and Daniel Junghans$^1$}\\

\vspace{0.5cm}

$^{1}${\it Institute for Theoretical Physics, University of Heidelberg, Philosophenweg 19,\\[.1cm]
 D-69120 Heidelberg, Germany}\\
 $^{2}${\it College of Physics, Sichuan University, Chengdu, 610065, China}

\vspace*{1cm}

\large{September 8, 2020}

\vspace{1.5cm}

\begin{abstract}

\noindent We analyze to which extent the KKLT proposal for the construction of de Sitter vacua in string theory is quantitatively controlled. Our focus is on the quality of the 10d supergravity approximation. As our main finding, we uncover and quantify an issue which one may want to call the ``singular-bulk problem''. In particular, we show that, requiring the curvature to be small in the conifold region, one is generically forced into a regime where the warp factor becomes negative in a significant part of the Calabi-Yau orientifold. This implies true singularities, independent of the familiar, string-theoretically controlled singularities of this type in the vicinity of O-planes. We also discuss possible escape routes as well as other control issues, related to the need for a large tadpole and hence for a complicated topology.

\end{abstract}

\end{center}

\newpage

\section{Introduction}

String compactifications to meta-stable de Sitter vacua are a key ingredient in string phenomenology as developed over the past 20 years. The leading candidates are the KKLT \cite{Kachru:2003aw} and LVS \cite{Balasubramanian:2005zx} proposals, which are however both not fully explicit. It is hence mandatory to keep questioning both the concrete proposals as well as the existence of stringy dS vacua in principle \cite{Danielsson:2018ztv, Ooguri:2018wrx} (see also \cite{Obied:2018sgi, Garg:2018reu}).
One way forward is to attempt to make the models fully explicit or to identify
problems that are hidden in some part of the constructions.

We will focus on KKLT as the earliest and simplest proposal (see also \cite{Akrami:2018ylq} for a review).
In the last years, various aspects of this construction have been scrutinized.
A series of papers studied the $\overline{\text{D3}}$-brane backreaction and revealed the appearance of flux singularities at the bottom of the warped throat \cite{Bena:2009xk, McGuirk:2009xx, Dymarsky:2011pm, Bena:2011hz, Bena:2011wh, Bena:2012bk, Gautason:2013zw, Blaback:2014tfa}. There are indications that these singularities are resolved by string theory \cite{Michel:2014lva, Cohen-Maldonado:2015ssa, Armas:2018rsy} (see also \cite{Blaback:2012nf, Bena:2012ek, Bena:2014jaa, Danielsson:2014yga, Bena:2016fqp} for a more pessimistic view), although an explicit regular solution remains to be constructed.
A more recent debate concerned a perceived problem with the 10d consistency of the non-perturbative effects that are necessary to stabilize the Kahler moduli \cite{Moritz:2017xto, Gautason:2018gln, Gautason:2019jwq}. As clarified in \cite{Hamada:2018qef, Kallosh:2019oxv, Hamada:2019ack, Carta:2019rhx, Kachru:2019dvo, Bena:2019mte, Grana:2020hyu}, this is not an issue if the moduli dependence of the gaugino condensate and a crucial four-gaugino term in the D7-brane action are correctly taken into account (see however \cite{Gautason:2019jwq}).\footnote{For
earlier work on non-perturbative effects in 10d see e.g.~\cite{Camara:2004jj, Koerber:2007xk, Baumann:2010sx, Heidenreich:2010ad, Dymarsky:2010mf}.
}
Another criticism is related to using effective field theory in flux compactifications with runaway potentials \cite{Sethi:2017phn} (see however \cite{Kachru:2018aqn}). Furthermore, several recent papers argued that the stabilization of the conifold modulus can be problematic \cite{Bena:2018fqc, Blumenhagen:2019qcg, Bena:2019sxm, Dudas:2019pls, Randall:2019ent}.

In this paper, we will be concerned with a different issue, which we call the \emph{singular-bulk problem}. In particular, we argue that KKLT moduli stabilization generically implies an inequality for the warp factor $h$ of the form
\begin{equation}
\frac{|\tilde{\partial h}|}{h} \gtrsim g_sM^2 \label{sw}
\end{equation}
near the 4-cycle that supports the non-perturbative effect (due to an E3 instanton or a D7-brane stack with a gaugino condensate). Here, $M$ is the quantized $F_3$ flux through the $S^3$ at the bottom of the warped throat. Since $g_sM\gtrsim 1$ is required for the curvature there to be small \cite{Klebanov:2000hb} and $M \gtrsim 12$ for meta-stability \cite{Kachru:2002gs}, the right-hand side of the inequality is a large number. As we will show, this generically implies that
the warp factor becomes singular over a large region of the original Calabi-Yau.

Our work is partly inspired by the paper \cite{Carta:2019rhx}, which also discusses strong warping in KKLT. However, our conclusions are rather different. In particular, it was argued in \cite{Carta:2019rhx} that the size of the warped throat in KKLT models with $h^{1,1}=1$  is too large to ``fit'' into the internal manifold. This was interpreted as an inconsistency of the geometry. However, as we will explain in more detail below, a large throat is by itself not an issue, as the supergravity equations guarantee the existence of a solution even when there is no clear distinction between a throat region and a weakly-warped bulk. The resulting geometry is strongly warped everywhere but may a priori still be well-defined and under control. On the other hand, the inequality we find is a much more severe problem as it implies that large parts of the geometry become singular. The threat of large singular regions was also discussed in the Appendix of \cite{Carta:2019rhx}, but without turning this into a quantitative problem for KKLT.

The problem we describe persists in several variants and generalizations of the original KKLT scenario, for example, allowing $h^{1,1} > 1$ and a potential generated by $F$-terms and $D$-terms, if the Kahler moduli are stabilized by non-perturbative effects. However, we find that the problem is ameliorated if $\alpha^\prime$ corrections are large enough to participate in the stabilization, as it is the case in the large-volume scenario \cite{Balasubramanian:2005zx}.

Independently of the singular-bulk problem, we also discuss a further problem, which is related to the requirement of a large D3 tadpole $N\gg 1$.
In this regime, the volume of the 4-cycle wrapped by the instanton is large such that string corrections seem to be well-controlled at first sight.
However, we give an argument suggesting that large $N$ may in fact lead to a loss of control. Intuitively, the problem is that increasing $N$ does not correspond to taking the usual large-volume limit. Instead, it means increasing the instanton volume while at the same time increasing some of the Hodge numbers. The topology of the Calabi-Yau or of the submanifold wrapped by a D7 brane therefore becomes more and more complicated in this limit. If the Hodge numbers grow fast enough with $N$, some of the cycle volumes can become sub-stringy even though the instanton volume and the Calabi-Yau volume are large. While we do not present a fully general analysis of this behavior in this work, we indeed confirm our claim explicitly under certain assumptions.

This work is organized as follows. In Sect.~\ref{sec:setup}, we review how stabilizing the Kahler moduli as in KKLT implies strong warping and argue that this is by itself not a problem for the consistency of the compactification.
In Sect.~\ref{sec:problem}, we derive the inequality \eqref{sw} and discuss the resulting singular-bulk problem. In Sect.~\ref{sec:escapes}, we analyze several escape routes and argue that most of them are likely to fail. In Sect.~\ref{sec:tadpole}, we study a further potential problem related to the requirement of a large tadpole. We conclude in Sect.~\ref{sec:concl} with a discussion of our results.
\\

\section{Throat radius vs.~Calabi-Yau volume}
\label{sec:setup}

In type-IIB orientifold compactifications with O3/O7 planes, the KKLT scenario \cite{Kachru:2003aw} proposes to stabilize all moduli
in a two-step procedure. First, all complex-structure moduli and the axio-dilaton are stabilized by the $F$-term conditions derived from the Gukov-Vafa-Witten (GVW) flux superpotential \cite{Gukov:1999ya, Giddings:2001yu}.
Second, the Kahler moduli are stabilized using non-perturbative effects coming from gaugino condensation or Euclidean D3-brane (E3-brane) instantons. In the simplest case of a single Kahler modulus $T$ and denoting the vacuum value of the flux superpotential by $W_0$, one has a 4d supergravity model based on
\begin{equation}
K=-3\log(T+\bar{T})\, ,\qquad W=W_0+A e^{-2 \pi T/N_C}\,.
\end{equation}
Here $N_C$ characterizes the gauge group of the $SU(N_C)$ gaugino condensation or $N_C\equiv 1$ for E3 instantons. A key point for us is that the resulting SUSY AdS minimum has vacuum energy
\begin{equation}
V_\text{AdS}\sim -e^{-4 \pi \mathrm{Re}(T)/N_C}\,,
\end{equation}
where we disregarded the non-exponential prefactor. The vacuum value of $\mathrm{Re}(T)$ relevant in the above is $\mathrm{Re}(T) \sim N_C/(2 \pi)\, {\rm log}\, |{W_0}|^{-1}$, with $|W_0|\ll 1$ (see \cite{Demirtas:2019sip} for recent progress on finding flux vacua with exponentially small $|W_0|$).

The critical third step is to uplift this to a de Sitter vacuum by adding an $\overline{\text{D}3}$ brane to a Klebanov-Strassler-type throat \cite{Klebanov:2000hb} assumed to be present in the geometry. The strong warping suppression at the bottom of this KS throat \cite{Giddings:2001yu} makes the uplift energy density parametrically small:
\begin{equation}
V_\text{uplift} \sim e^{-8\pi K/ 3g_sM}\,.
\end{equation}
Here $K$ and $M$ are the flux numbers of the two 3-cycles characterizing the throat and we again disregarded all non-exponential effects.

Crucially, meta-stable de Sitter vacua arise only if
$|V_\text{AdS}|\sim |V_\text{uplift}|$ and hence
\begin{equation}
\mathrm{Re}(T)\simeq \frac{2N_C K}{3 g_s M}\label{tkm}
\end{equation}
is required. As described in \cite{Carta:2019rhx}, this may lead to problems associated with the necessity to `glue' the KS throat into the compact Calabi-Yau. Indeed, the 3-form fluxes characterized by $K$ and $M$ induce a D$3$ tadpole $N\equiv KM$ localized in the throat region.
Such a localized tadpole leads to strong warping within a (string-frame) distance set by
\begin{equation}
R_\text{throat}^4 \simeq 8\pi g_s N \alpha'^2= 4g_sN/(2\pi)^3\,.
\label{rth}
\end{equation}
This is well known from the geometry of a throat sourced by a D$3$-brane stack \cite{Verlinde:1999fy}. Here in the last step (and in what follows) we use the standard convention of setting $\ell_s\equiv 2\pi\sqrt{\alpha'}=1$.

Now, for the throat to be glued in a weakly-warped Calabi-Yau, one requires $R_\text{throat}\lesssim R_\text{CY}$. Defining
a typical string-frame Calabi-Yau radius by $R_\text{CY}^4\sim g_s \mathrm{Re}(T)$ and dropping the $(2\pi)$ factors in (\ref{rth}), the condition $R_\text{throat}\lesssim R_\text{CY}$ becomes
\begin{equation}
g_s N\lesssim \frac{N_C K}{M}\,.\label{gsn}
\end{equation}
Focussing on instantons (i.e. $N_C=1$)\,\footnote{
It was argued in \cite{Carta:2019rhx} that large $N_C$ does not improve the situation due to the complicated topology required to cancel the large D$7$ tadpole.
}
and using $K=N/M$, this implies
\begin{equation}
{\cal O}(1) \lesssim \frac{1}{g_s M^2}\lesssim \frac{1}{M}\,.\label{mmm}
\end{equation}
Here the second inequality uses $g_sM\gtrsim 1$, required by the supergravity approximation at the bottom of the throat. Finally, one needs $M\gtrsim 12$ by stability against brane-flux annihilation \cite{Kachru:2002gs}, leading to an apparent contradiction. Different bounds on $M$, with similar effects in our context, have been suggested in \cite{Bena:2018fqc, Blumenhagen:2019qcg, Bena:2019sxm, Dudas:2019pls, Randall:2019ent}.
An attempt to collect the $(2\pi)$ and other prefactors and hence to evaluate the numerical severity of the problem is presented in Appendix~\ref{nums}. Moreover, if one is willing to view $M$ and $g_sM$ as parameters which need to be large to ensure meta-stability and the validity of supergravity, then the inequality in (\ref{mmm}) represents a parametric rather than just a numerical problem.

Let us be more precise about how the assumed bound $M\gtrsim 12$ arises. While a decay via brane polarization as discussed in \cite{Kachru:2002gs} requires $p>1$ $\overline{\text{D}3}$ branes, it is also clear that a single $\overline{\text{D}3}$ brane cannot be absolutely stable but must be able to decay into a supersymmetric vacuum. This decay involves a change in the flux numbers, so it must happen in a non-local way via a domain wall. One therefore expects that an NS5 brane carrying the $\overline{\text{D}3}$ charge can nucleate and potentially cause an instability for any $p$ \cite{Kachru:2002gs, Michel:2014lva}. Whether this NS5 brane is created by brane polarization or some other (stringy) process is irrelevant in this context. In any case, in order to prevent an instability via such an NS5 brane slipping over the $S^3$ at the tip of the KS throat, the NS5-brane potential needs to have a barrier. Since the region of the potential near the maximum lies at a large radius of the order $\sqrt{g_sM}$, we expect that the bound $M\approx 12p$ obtained in \cite{Kachru:2002gs} for the vanishing of the barrier should also apply to the case $p=1$.\footnote{Properly accounting for the NS5-brane backreaction may increase the numerical prefactor in the bound on $M$ \cite{Kachru:2002gs}.}
An even stronger bound $g_sM^2 \gtrsim (6.8\sqrt{p})^2$ was argued in \cite{Bena:2018fqc, Blumenhagen:2019qcg} to be required (for any $p$) in order to prevent a runaway of the conifold modulus to zero. The derivation assumes a specific off-shell field dependence of
the warp factor at the $\overline{\text{D}3}$ position in a regime where the conifold modulus deviates by a large factor from its flux-stabilized value. It is not clear to us whether this is justified. In the remainder of this paper, we will work with the more conservative assumption that $g_sM^2 \approx 12$ demarcates the boundary of control.

While the set of arguments leading to \eqref{mmm} is very inspiring, it is not obvious to us that this constitutes a technically well-defined problem for KKLT. To see this, let us for the moment turn off $\overline{\text{D}3}$ and non-perturbative effects. This takes our discussion back to the level of GKP \cite{Giddings:2001yu}, where the volume modulus $T$ is a flat direction. Now we can go to the specific value of $\mathrm{Re}(T)$ where the throat `just' fits into the weakly-warped Calabi-Yau space. The arguments above tell us that, in consistent KKLT setups, $\mathrm{Re}(T)$ tends to be stabilized at a smaller value. But for the moment $\mathrm{Re}(T)$ is a flat direction and we are allowed to go to smaller $\mathrm{Re}(T)$. This is guaranteed by the equations of motion, which admit a solution for all values of $\mathrm{Re}(T)$, even though the resulting manifold cannot be visualized as a throat glued to a weakly warped bulk anymore.\footnote{Although in fact \cite{Carta:2019rhx} presented such a picture, dismissing it as very non-generic.
}
Instead, we will have a significantly varying warp factor also outside the KS
throat.
However, it is a priori not clear why this would be problematic.

In particular, even when the warp factor is not approximately constant, it may still be positive over most of the compactification space (except near the familiar O-plane singularities). Strongly-warped flux compactifications can therefore very well be trustworthy supergravity solutions. Although the appendix of \cite{Carta:2019rhx} displays a suggestive figure with a large singularity, it is not immediately clear how to turn this into an argument against KKLT. Indeed, as one enters the strong-warping regime, the warp factor is shifted by a negative constant \cite{Giddings:2001yu, Giddings:2005ff} so that any negative region of this function certainly grows. But the non-trivial task is to show quantitatively that a large singularity appears at the point corresponding to the KKLT minimum. Moreover, it is essential to achieve this in the naively most benign setting where the negative tadpole is distributed in a generic way on the Calabi-Yau.

In what follows, we analyze precisely this question. We will argue that there is indeed a technical problem.

\section{The singular-bulk problem}
\label{sec:problem}

In this section, we formulate the singular-bulk problem. For concreteness, we assume a weakly coupled type IIB description, i.e., the dilaton satisfies $g_s\lesssim 1$ and is approximately constant over the compact space. This is true for models with O3 planes and for models with 7-branes in the orientifold limit (except very close to the 7-branes). We furthermore assume $h^{1,1}_+=1$ as in the original KKLT scenario. Generalizations to $h^{1,1}_+>1$ will be discussed in Sect.~\ref{sec:escapes}.

\subsection{Basic argument}
\label{sec:basic}

Under the above assumptions, the 10d string-frame metric takes the form \cite{Giddings:2001yu}
\begin{equation}
\d s_{10}^2 = h(y)^{-1/2} \eta_{\mu\nu}\d x^\mu \d x^\nu + h(y)^{1/2} \tilde g_{mn}\d y^m \d y^n,
\end{equation}
where $h$ is the warp factor and $\tilde g_{mn}$ is a 6d metric which is Ricci-flat. Without loss of generality, we normalize $\tilde g_{mn}$ such that ${\cal \tilde V}\equiv\int_X \sqrt{\tilde g} = 1$.

As discussed in the previous section, a successful uplift requires a certain relation between the warp factor in the throat and the volume of the 4-cycle (let us call it $\Sigma$) on which the E3 brane or the D7-brane stack is wrapped. Concretely, this relation was given in (\ref{tkm}) and we may write it as
\begin{equation}
g_s\mathrm{Re}(T)\simeq \frac{2N}{3 M^2}\,, \label{ret}
\end{equation}
where we have focused on the E3 case and hence $N_C=1$ for simplicity. We also recall that the expression
$2\pi \mathrm{Re}(T)$ in the exponent of the non-perturbative superpotential is precisely the E$3$ volume $V_\Sigma$ multiplied with the D$3$-brane tension $2\pi/g_s$.\footnote{The precise definition of the Kahler coordinates in warped flux backgrounds is subtle (see \cite{Martucci:2016pzt} and references therein). However,
our interest is in the exponent of the suppression factor of the instanton effect, which we expect to be quantified by the DBI action of the E3 brane. Further note that E3 branes may carry a non-zero worldvolume flux $\mathcal{F}$ (see, e.g., \cite{Blumenhagen:2007bn, Grimm:2011dj, Bianchi:2011qh, Bianchi:2012kt}). We neglect this possibility here for simplicity. However, one can verify (using $\sqrt{g|_\Sigma+\mathcal{F}}\approx \sqrt{g|_\Sigma}\,(1+\frac{1}{2}|\mathcal{F}|^2)$) that the flux modifies the DBI action in such a way that the bounds on $h$ derived below become even stronger.
We assume further $\alpha^\prime$ corrections to be negligible at large volume (see Sect.~\ref{sec:curv} for a possible exception).
} Thus, we also have
\begin{equation}
V_\Sigma \simeq \frac{2N}{3 M^2} \qquad \mbox{or}\qquad
\langle h \rangle_\Sigma \simeq \frac{2N}{3 \tilde V_\Sigma M^2}\,\,. \label{mod}
\end{equation}
Here $\tilde V_\Sigma \equiv \int_{\Sigma} \sqrt{\tilde g|_\Sigma}$ is the string-frame 4-cycle volume as measured with the tilded metric and $\langle h \rangle_\Sigma \equiv  \int_{\Sigma} \sqrt{\tilde g|_\Sigma}\,h/\tilde V_\Sigma=V_\Sigma/\tilde V_\Sigma$ is the warp factor averaged over that 4-cycle.\footnote{The reader may want to recall that, in this parameterization, the volume modulus is encoded in the warp factor. More precisely, changing the volume Re$(T)$ corresponds to a constant shift $h\to h+\mbox{const.}$}

Note that, generically, $\tilde V_\Sigma\sim \mathcal{O}(1)$ due to our normalization of $\tilde g_{mn}$. In fact, using the string-frame 2-cycle volume $\tilde t$, one has ${\cal \tilde V}=\kappa_{111}\tilde t^3/3!$ and $\tilde V_\Sigma \ge \partial{\cal \tilde V}/\partial \tilde t = \kappa_{111}\tilde t^2/2$, with an integer triple-intersection number $\kappa_{111}$.\footnote{The inequality symbol is due to the fact that $\Sigma$ is not necessarily the minimal-volume cycle in $\tilde g$ but only in the conformally-Calabi-Yau metric $g$.} This implies $\tilde V_\Sigma \ge \kappa_{111}^{1/3}(6{\cal \tilde V})^{2/3}/2$, such that even if models with large $\kappa_{111}$ can be found, $\tilde{V}_\Sigma$ will become larger than unity and our problem will become more severe.

The above implies that there is a neighborhood of a point $y_0$ on $\Sigma$ for which
\begin{equation}
h \lesssim \frac{2N}{3 \tilde{V}_\Sigma M^2}. \label{ineq0}
\end{equation}
This is problematic since the backreaction of the flux localized in the conifold causes the variation of the warp factor in the bulk to be much larger than $N/M^2$. To see this, recall that the warp factor obeys a differential equation similar to the electrostatic potential on a compact space, with a collection of positive and negative electric charges \cite{Giddings:2001yu, Giddings:2005ff}:
\begin{equation}
-\tilde{\nabla}^2\,h\,= 2 g_s^2 \kappa_{10}^2 T_3 \,\tilde{\rho}_{\text{D}3} = g_s\, \tilde{\rho}_{\text{D}3}\,.\label{epp}
\end{equation}
The simplification of the prefactor in the last expression relies on our conventions for $\ell_s$. The D3-charge density $\tilde{\rho}_{\text{D}3}$ is defined such that a single D3 brane contributes a $\delta$-function in the $\tilde{g}$ metric.\footnote{
Here
and below, $\tilde \nabla$ denotes the covariant derivative adapted to $\tilde g$. In the flat limit, our Laplace equation is consistent with the familiar solution $h\simeq 4\pi\alpha'^2 g_s N/r^4$ in the vicinity of a stack of $N$ D3 branes.
}
In our context, $N$ units of positive D$3$ charge are supplied by the 3-form flux near the 3-cycle at the tip of the deformed conifold. In the simplest case, we may neglect further positive contributions from flux elsewhere in the Calabi-Yau. The compensating negative contribution comes from O3 planes scattered throughout the Calabi-Yau and/or curved 7-branes. (We will use the picture of scattered O3 planes for simplicity, although the required large tadpole might equally well come from the curvature of 7-branes. The distinction is not important for our purposes.)

As a result, $h$ is the solution to an electrostatic-potential problem as defined by \eqref{epp}, with charge $g_sN$ at one point of the Calabi-Yau and the compensating background charge scattered through the rest of the space. The size of the space is ${\cal O}(1)$ by our normalization of $\tilde{g}$. Thus, for lack of any large or small parameter, the solution $h$ will typically vary by an ${\cal O}(1)$ amount on an ${\cal O}(1)$ distance scale if $g_sN=1$. But for us,  $g_sN$ is our central large parameter. We therefore find
\begin{equation}
|\tilde{\partial h}| \sim g_sN \label{var}
\end{equation}
for a generic point in the Calabi-Yau (including points near $\Sigma$), where we define $|\tilde{\partial h}|\equiv \sqrt{\tilde g^{mn}(\partial_m h) (\partial_n h)}$ as our proxy for how strongly $h$ varies. Alternatively, one may derive the scaling \eqref{var} by evaluating at an $\mathcal{O}(1)$ distance the explicitly known warp factor of the deformed conifold \cite{Klebanov:2000hb, Herzog:2001xk}.

Combining \eqref{ineq0} and \eqref{var}, we have a neighborhood of a point $y_0$ on $\Sigma$ where
\begin{equation}
\boxed{\frac{|\tilde{\partial h}|}{h} \gtrsim g_sM^2 \gtrsim M \gg 1.} \label{ineq}
\end{equation}
Here we used that $M$ must be a fairly large number, as discussed in Sect.~\ref{sec:setup}. Since $h(y_0+\delta y) \approx h(y_0)+\partial_m h(y_0) \delta y^m$, it follows that we generically run into a singularity $h= 0$ at a distance $|\tilde{\delta y}| \lesssim 1/(g_sM^2) \ll 1$ (as measured with $\tilde g$). We call this the {\bf singular-bulk problem}. The problem is parametric rather than just numerical if we are prepared to think of $g_sM\gtrsim 1$ \cite{Klebanov:2000hb} or of the bound $M_\text{min}\simeq 12$ of \cite{Kachru:2002gs} as of large parameters.

Crucially, the singularities implied by the above argument are independent of the usual singularities in the vicinity of O-planes. The latter are believed to be resolved by string theory and hence to not represent a problem. Our claim is rather that the singularity we find is created by the too strong variation of $h$ due to the positive D3 charge in the KS throat. We will give detailed arguments for this interpretation in Sect.~\ref{sec:sing}.

Let us close the current section with two remarks.
First, note that an alternative perspective on the singular-bulk problem is to consider the internal curvature scalar. Using the Ricci-flatness of $\tilde{g}$, it takes the form
$R_6 = h^{-5/2} |\tilde{\partial h}|^2 - \frac{3}{2} h^{-3/2} \tilde \nabla^2 h$ in the string frame. According to the warp factor equation \eqref{epp}, $\tilde\nabla^2 h \le 0$ at any point away from an O3 plane. Employing \eqref{ineq0} and \eqref{var}, we therefore find $R_6 \gtrsim g_s^2M^5/\sqrt{N}$. To ensure meta-stability of the $\overline{\text{D}3}$ brane and  control over $\alpha'$ corrections, we require $M\gtrsim 12$, $g_sM\gtrsim 1$ and $R_6 \lesssim 1$. Hence, even at the boundary of control, the tadpole would have to be of the order $N \gtrsim g_s^4M^{10} \gtrsim 12^{6}\approx 3\cdot 10^6$. This significantly exceeds the largest known tadpole $N\approx 7.5\cdot 10^4$ in a type IIB/F-theory compactification \cite{Taylor:2015xtz}.
We thus again conclude that the supergravity solution breaks down in the vicinity of $\Sigma$.

As a final remark,
note that, for a completely general function, \eqref{ineq} would not necessarily imply a nearby root.
However, as explained above, $h$ is not a general function but determined by the Poisson equation \eqref{epp} with a source of charge $g_sN$. It is therefore expected to take the form $h=g_sNh_0 + \ldots$ in the bulk, where the dots are local contributions from O-planes and $h_0$ is an $\mathcal{O}(1)$ function, i.e., $h_0$ does not depend on any large or small parameters and thus varies by an $\mathcal{O}(1)$ amount over an $\mathcal{O}(1)$ distance. In other words, $h$ generically has $\mathcal{O}(g_sN)$ Taylor coefficients. For $|\tilde{\delta y}| \ll 1$, it is then justified to neglect quadratic and higher orders in the Taylor expansion as we have done above and the singularity follows.
This logic only fails if $y_0$ happens to be an \emph{ungeneric} point near or at an extremum where the first-order Taylor coefficient $\partial_m h$ vanishes and \eqref{ineq} is violated (recall that the argument leading to \eqref{ineq} assumed $y_0$ to be generic). However, as we will now show, a singularity can be inferred even in this special case.

Since the variation of $h$ is $\mathcal{O}(g_sN)$, we expect that higher derivatives are still of the order $g_sN$ at such a point. In particular, we generically have $\partial_m\partial_n h \sim g_sN$ (in an orthonormal frame).\footnote{More generally, one could also consider a point at which not only the first derivative but the first $n$ derivatives of $h$ vanish. We will not study this possibility as it is even more ungeneric than the case discussed here. Note, however, that arguments involving the $(n+1)$th derivative of $h$ may then still lead to similar conclusions.
} Because of $\partial_m h = 0$, this implies $\tilde\nabla_m\partial_n h \sim g_sN$. However, the warp factor equation \eqref{epp} states that $\tilde\nabla^2 h \le 0$ at any point (away from an O3 plane).
Therefore, $\tilde\nabla_m\partial_n h$ must have at least one negative eigenvalue, i.e., we expect that $\text{min} \big( \tilde\nabla_m \partial_n h \big) \sim -g_sN$ at $y_0$. Dividing by \eqref{ineq0}, we can write this as the condition
\begin{equation}
\frac{\text{min} \big( \tilde\nabla_m \partial_n h \big)}{h} \lesssim - g_sM^2 \label{ineq2}
\end{equation}
in an orthonormal frame. As before, this implies again a small $|\tilde{\delta y}|$ for which $h(y_0+\delta y) \approx h(y_0)+\tilde\nabla_m\partial_n h(y_0) \delta y^m\delta y^n/2$ is zero.

\subsection{A closer look at the singularity}
\label{sec:sing}

In this section, we study in more detail the singularity found above. As stated before, we claim that it cannot be identified with the usual O-plane singularities that should be ok in string theory. It is instead created by the large variation of the warp factor due to the D3 charge in the KS throat and indicates a pathology of the compactification. To see the distinction, we give three arguments:

First, our singularity covers a large part of the Calabi-Yau including almost the whole E3 volume, whereas the singular region surrounding an O3 plane would be a small local effect.\footnote{Note that volumes and distances cannot be meaningfully computed in the physical metric $g$ in a singular region since the metric formally becomes imaginary there. It is however a well-defined question to ask for the corresponding volumes/distances in $\tilde g$, which is smooth and positive everywhere.} To see this, assume that the E3 volume is dominated by $h>0$ regions where supergravity is valid (with O-plane singularities, if present in the vicinity of the E3, resolved by string theory such that they are negligible in the relevant integral). This leads to a contradiction because the function $h$ can then only take generic values $h\sim g_sN$ on a small fraction $\lesssim 1/g_sM^2$ of the E3 volume (as measured with $\tilde g$) or else \eqref{mod} cannot be satisfied. Everywhere else on the E3, $h$ must be smaller than $N/M^2$. This means that we can repeat our above singularity argument at almost every point on the E3,
thus contradicting our initial assumption. We thus find that the singularity must extend over an $\mathcal{O}(1)$ fraction of the E3 volume. This generically implies that the negative region also spreads over an $\mathcal{O}(1)$ distance into the transverse space.\footnote{Here, by ``generic'' we mean that the negative region is not unnaturally thin, i.e., that near the E3 the variation of $h$ along the transverse directions is not much larger than its variation along the worldvolume directions.} We thus conclude that the singularity covers a large part of the Calabi-Yau.
This is to be contrasted with an O3 singularity: Imagine a thought experiment where we consider a 4-cycle with smooth $h\sim N/M^2$ and then add a negative O3 source to the warp-factor equation \eqref{epp}.\footnote{To be consistent with tadpole cancelation, we then also have to increase the D3 charge in the KS throat by the same amount. The effect of this is locally negligible near the O3 source.} 
The exact solution for the warp factor is of course not known on a Calabi-Yau, but locally the space is just $\mathbb{R}^6$ such that the new source should create a $1/r^4$ singularity (with $r=|y\,\widetilde{\!-y\,}\!_\text{O3}|$), as expected for a codimension-6 object. The new warp factor is therefore $h=h_\text{old}+h_\text{O3}$ with $h_\text{old}\sim N/M^2$ and $h_\text{O3}\sim - g_s/r^4+\mathcal{O}(1/r^3)$.
We thus see that the typical size of an O3 singularity (i.e., the value of $r$ at which $|h_\text{O3}| \sim h_\text{old}$) is formally $r\sim (g_s M^2/N)^{1/4} \ll 1$. This is parametrically smaller than the $\mathcal{O}(1)$ size of our singularity, which confirms our claim.\footnote{Of course, we do not trust supergravity in the vicinity of the singularities. The reader may therefore wonder about the significance of such estimates. However, it is clear that not \emph{every} singular solution to the supergravity equations corresponds to a well-defined string-theory background. It is therefore important to understand whether a singularity can be identified with a known object in string theory, even though the true physics near such an object is not described by supergravity. See also \cite{Cribiori:2019clo} for a similar criticism of a type IIA dS solution proposed in \cite{Cordova:2018dbb, Cordova:2019cvf}.
}

\begin{figure}[t!]
\centering
\includegraphics[trim = 90mm 50mm 90mm 40mm, clip, width=0.4\textwidth]{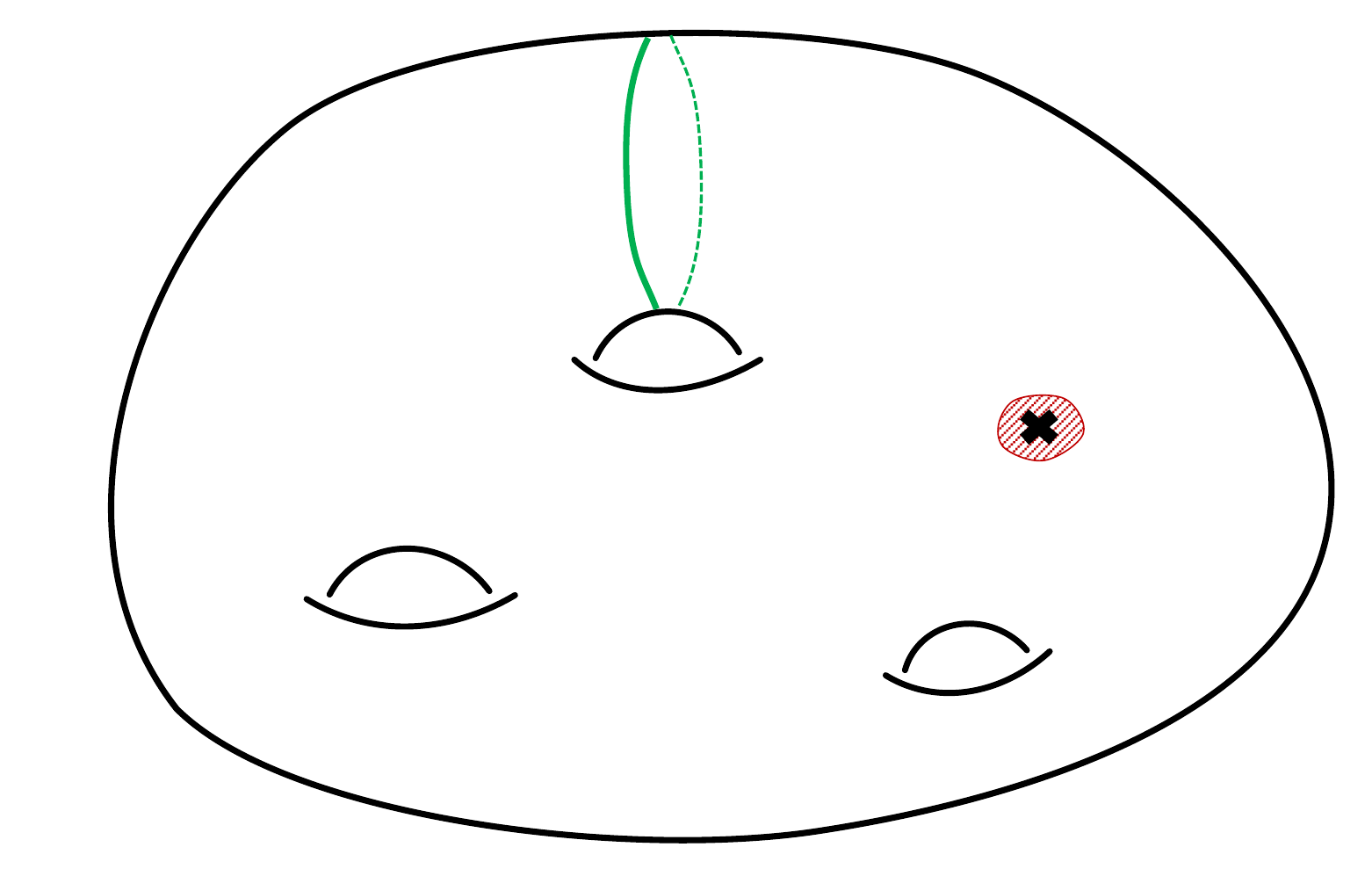} $\qquad\qquad$
\includegraphics[trim = 90mm 50mm 90mm 40mm, clip, width=0.4\textwidth]{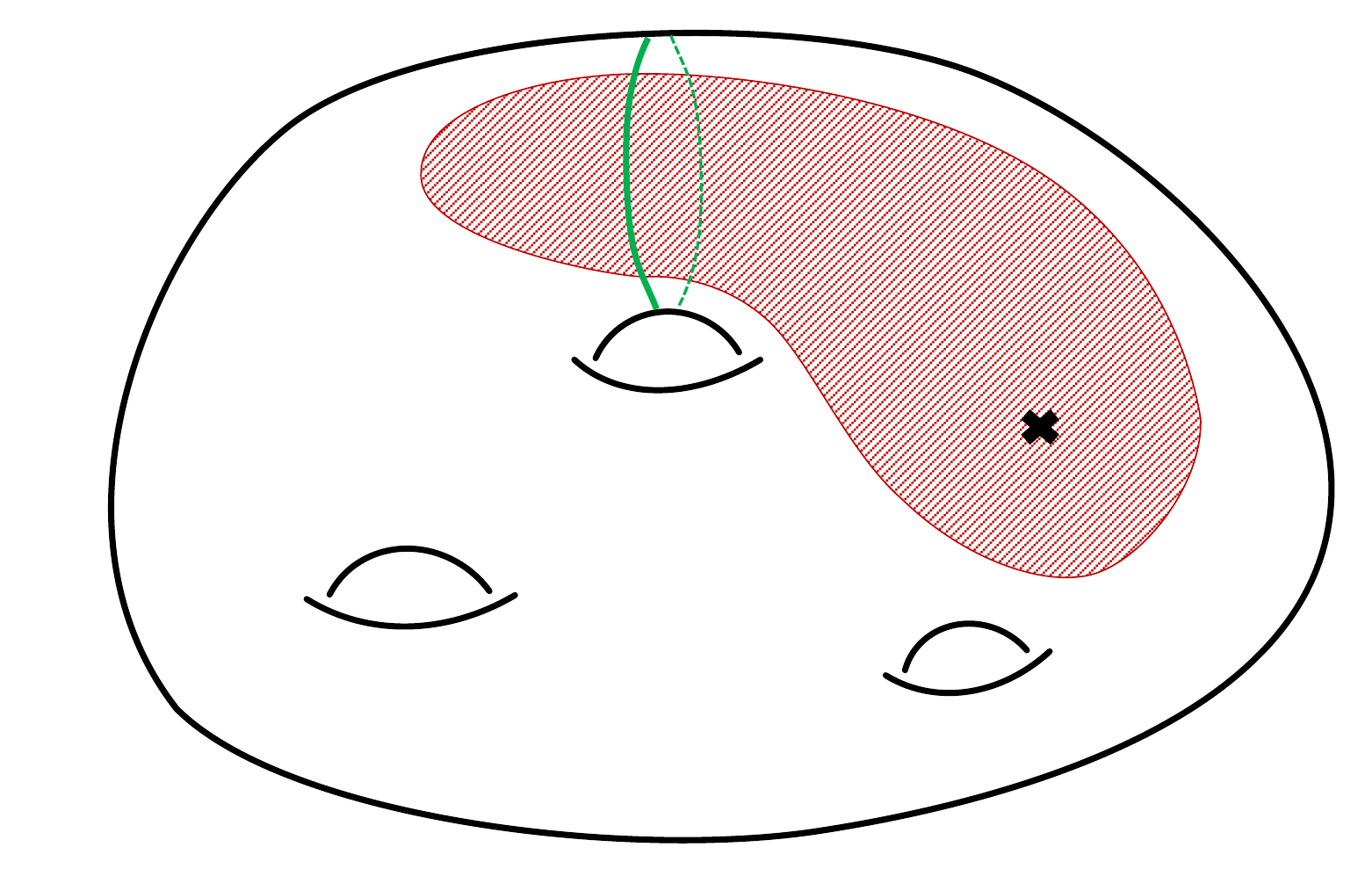}

\caption{Sketch of a Calabi-Yau orientifold with a 4-cycle supporting a non-perturbative effect (green) and singular regions (red). The left-hand side shows the expected singularity in the vicinity of an O-plane (black cross). The right-hand side shows the large singular region that appears in a flux compactification compatible with a KKLT-like uplift.
\label{sing}}
\end{figure}

Second, let us present an alternative argument that the singularity can extend over a large part of the Calabi-Yau. Consider a model as in Fig.~\ref{sing} where the nearest O3 plane is far away (i.e., at an $\mathcal{O}(1)$ distance in $\tilde g$) from the E3 location.
Our claim is that the singularity then extends from the E3 to (at least) the location of that O3 plane. To see this, recall that \eqref{epp} implies that at least one eigenvalue of $\tilde \nabla_m\partial_n h$ is negative in regions with net D3 charge (i.e., positive $\tilde\rho_{\text{D}3}$). Therefore, $h$ does not have a minimum anywhere on the Calabi-Yau. Now consider the 6d space surrounding the singularity, with boundary $h=0$. Since $h$ does not have a minimum, it must fall to minus infinity somewhere in this space, i.e., the space must contain at least one O-plane locus. In other words, there is a connected region on the Calabi-Yau for which $h$ stays negative all the way from $\Sigma$ until (at least) the nearest O-plane. Depending on where this O-plane sits in a particular model, large parts of the Calabi-Yau can disappear behind the singularity, see Fig.~\ref{sing}.

\begin{figure}[t]
\centering
\includegraphics[width=0.45\textwidth]{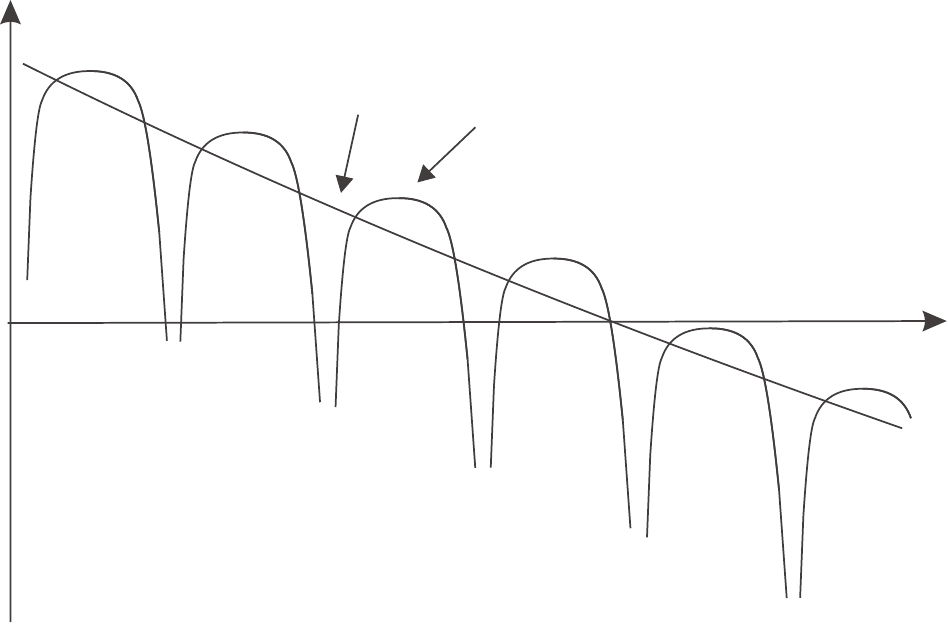}
\begin{picture}(0,0)
\put(0,60){$y$}
\put(-220,60){$0$}
\put(-220,130){$h$}
\put(-145,120){$h_c(y)$}
\put(-100,110){$h(y)$}
\end{picture}
\vskip-0.3cm
\caption{One-dimensional illustration of the behavior of the warping function $h(y)$, showing in particular the singularities at O3-plane loci. The corresponding coarse-grained function $h_c$ is also displayed. Its non-trivial long-distance profile, illustrated here as a negative overall tilt, is related to the global distribution of positive and negative charge in the Calabi-Yau.
\label{hc}}
\end{figure}
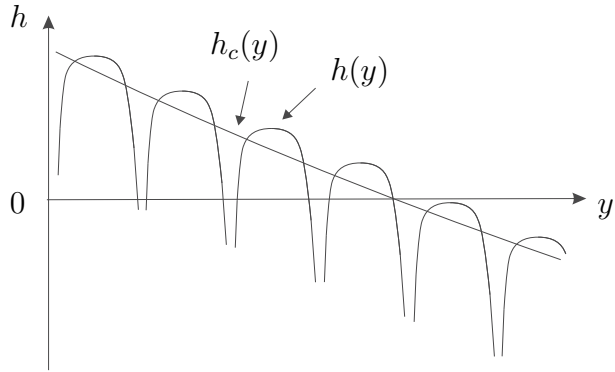

Third, the difference from an O3 divergence can be demonstrated using a {\it coarse-grained} warp factor. One may think of $h$ intuitively as of a potential in a plasma of electric charges. Clearly, such a potential becomes arbitrarily negative near negatively-charged point-like particles.
A one-dimensional illustration of the corresponding behavior of the function $h(y)$ is provided in Fig.~\ref{hc}. As also shown in the figure, there is in addition a physically meaningful averaged or coarse-grained potential $h_c(y)$. For the simple case of charges in flat space, $y\in \mathbb{R}^6$, a coarse graining of the charge distribution will result in an exactly analogous coarse graining of the induced potential. For example, using a Gaussian with width $d$, one has
\begin{equation}
h_c(y)=\frac{\displaystyle \int \d^6y'\,h(y')\, \exp(-|y-y'|^2/d^2) }{ \displaystyle \int \d^6y'\,\exp(-|y-y'|^2/d^2)} \,. \label{cgd}
\end{equation}
In our context, we should think of the flat $\mathbb{R}^6$ metric as the analogue of the unwarped Calabi-Yau metric $\tilde g$. We take the coarse-graining scale $d$ to be larger than the typical distance between O3 planes, $d\gg d_{\rm O3}$, and smaller than the distance to the throat.
This is consistent in the large-$N$ limit (neglecting tadpole constraints for the moment) where $d_{\rm O3}\ll 1$.
The resulting coarse-grained charge distribution is then a positively charged lump of size $d$ in the throat and a smooth distribution of negative charge in the bulk.

Crucially, as we tried to illustrate in Fig.~\ref{hc}, the coarse-grained function closely follows the maxima of the function $h(y)$. The reason is that, as stated before, the function $h$ near any O3 location $y_{\rm O3}$ behaves like
\begin{equation}
h(y)\sim -\frac{1}{|y-y_{\rm O3}|^4}\,.
\end{equation}
This is simply the standard behavior of an electrostatic potential near a negative charge in 6 spatial dimensions. As a result, near-O3 regions contribute to \eqref{cgd} like $\int \d^6x/|x|^4$, i.e., only insignificantly. Indeed, let $\epsilon \ll d_{\rm O3}$ define what we call the near-O3 region. Then, disregarding the exponential term in (\ref{cgd}) since it is irrelevant at short distances, the crucial estimate is
\begin{equation} \int\limits_{|y'-y_{\rm O3}|<\epsilon} \frac{\d^6y'}{|y'-y_{\rm O3}|^4}\,\,\,\ll \int\limits_{\epsilon<|y'-y_{\rm O3}|<d_{\rm O3}} \frac{\d^6y'}{|y'-y_{\rm O3}|^4}\,.\label{sri}
\end{equation}
Hence, as already stated above, the function $h_c$
follows the maxima of $h$ closely.

Now, we expect that the essential points of the above argument hold not only for a charge distribution in $\mathbb{R}^6$ but also in the Calabi-Yau. The basis of this expectation is that the overall logic of coarse graining can clearly be applied to both compact and non-compact spaces. Moreover, the estimate of the effect of singular regions in the averaging process given in (\ref{sri}) characterizes a short-distance phenomenon, where the Calabi-Yau geometry and topology do not matter.

With this, we return to our main line of reasoning. The key point is that our argument about
negative regions of $h(y)$, developed around (\ref{var}) and (\ref{ineq}), applies equally well to $h_c(y)$. In particular, as long as $d\ll 1$, the coarse-grained O3 charge does not significantly overlap with the coarse-grained positive charge in the KS throat. We therefore still have a positive charge $g_sN$ in the throat, up to negligible corrections. Our crucial estimate (\ref{var}) for the warp-factor variation therefore also holds for $h_c$. At the same time, it would be highly ungeneric if $h_c$ could be much larger than $\langle h \rangle_\Sigma$ everywhere on the E3.
It then follows from \eqref{mod} that $h_c$ satisfies \eqref{ineq0}. The rest of our argument then goes through as before. We thus conclude that the coarse-grained warp factor develops a singularity. Crucially, this happens even though the O3 charge is now distributed smoothly over the bulk and all negative ``spikes'' in $h$ corresponding to local O3 divergences are completely washed out by the averaging.

Furthermore, as argued in the beginning of this section, this is not a small effect: Indeed, a large part of the Calabi-Yau (in the $\tilde{g}$ metric) becomes singular in the physical string-frame metric $g$ that is defined on the basis of the (non-coarse-grained) warp factor $h$. Thus, observing from Fig.~\ref{hc} what happens when $h_c$ turns negative, we conclude that the negative region generically includes many O3 planes and presumably many of the 3-cycles and some part of the flux of the Calabi-Yau. This is very different from the small and presumably harmless singularities near the O3 planes as they are present, for example, on the l.h.~side of the sketch in Fig.~\ref{hc}.

\subsection{A toy model}
\label{sec:toymodel}

\begin{figure}[t]
\centering
\includegraphics[width=5cm]{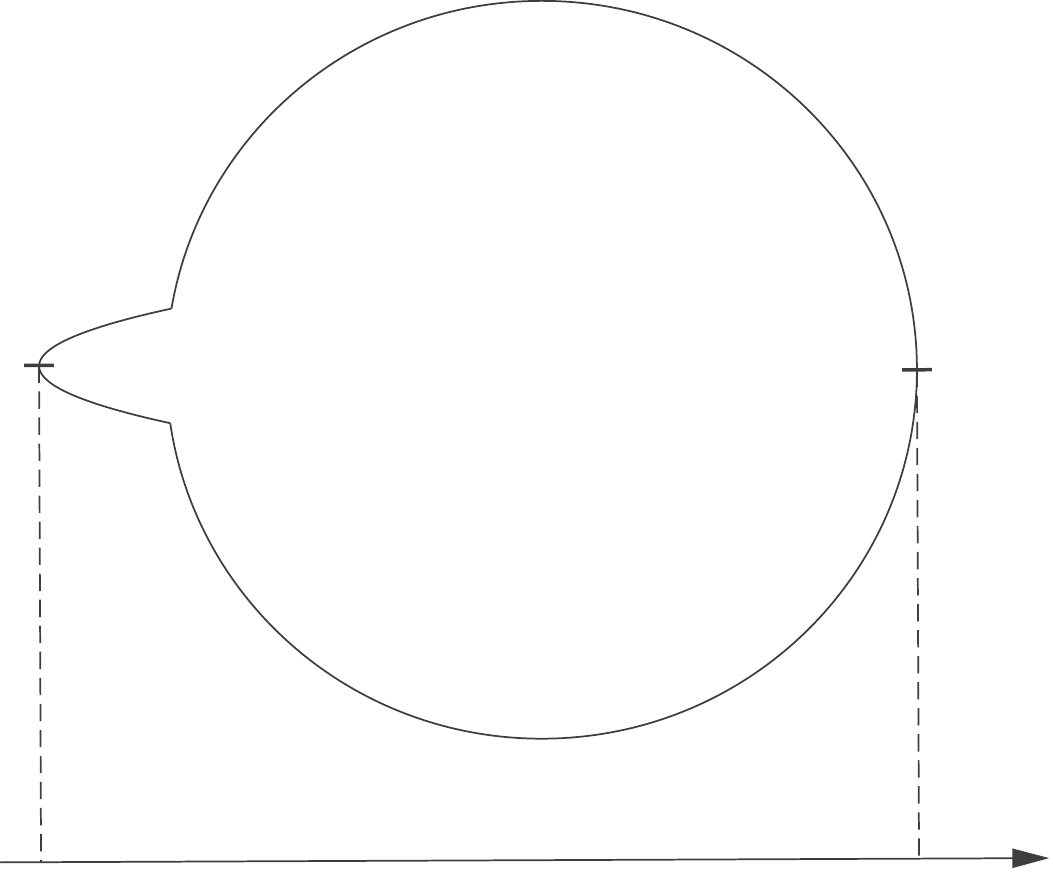}
\begin{picture}(0,0)
\put(-5,-10){$\phi$}
\put(-25,-10){$\pi$}
\put(-143,-10){$0$}
\put(-155,65){$\text{N}$}
\put(-15,65){$\text{S}$}
\end{picture}
\vspace{1em}
\caption{Compact space modeled as an $S^6$ with a `conifold region' glued in at the north pole N. The latitude on the sphere is parameterized by an angle $\phi$ as shown.\label{s6}}
\end{figure}
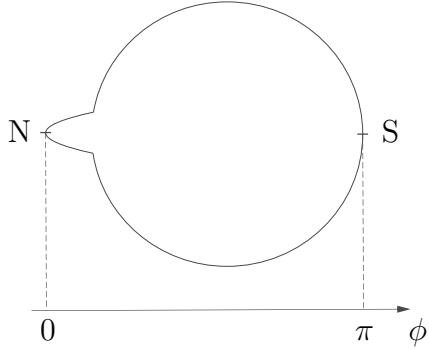

Let us illustrate our general finding with a simple toy model that captures the essentials of the problem. We model our compact space as a unit-radius $S^6$. The latter can be viewed as a fibration of an $S^5$ over an interval $\phi \in (0,\pi)$. The $S^5$ radii are given by $R_5(\phi) = \sin(\phi)$. The warped throat is modeled by a nearly point-like source with charge $N$ at the north pole $\phi=0$, cf.~Fig.~\ref{s6}.

\begin{figure}[t]
\centering
\includegraphics[width=12cm]{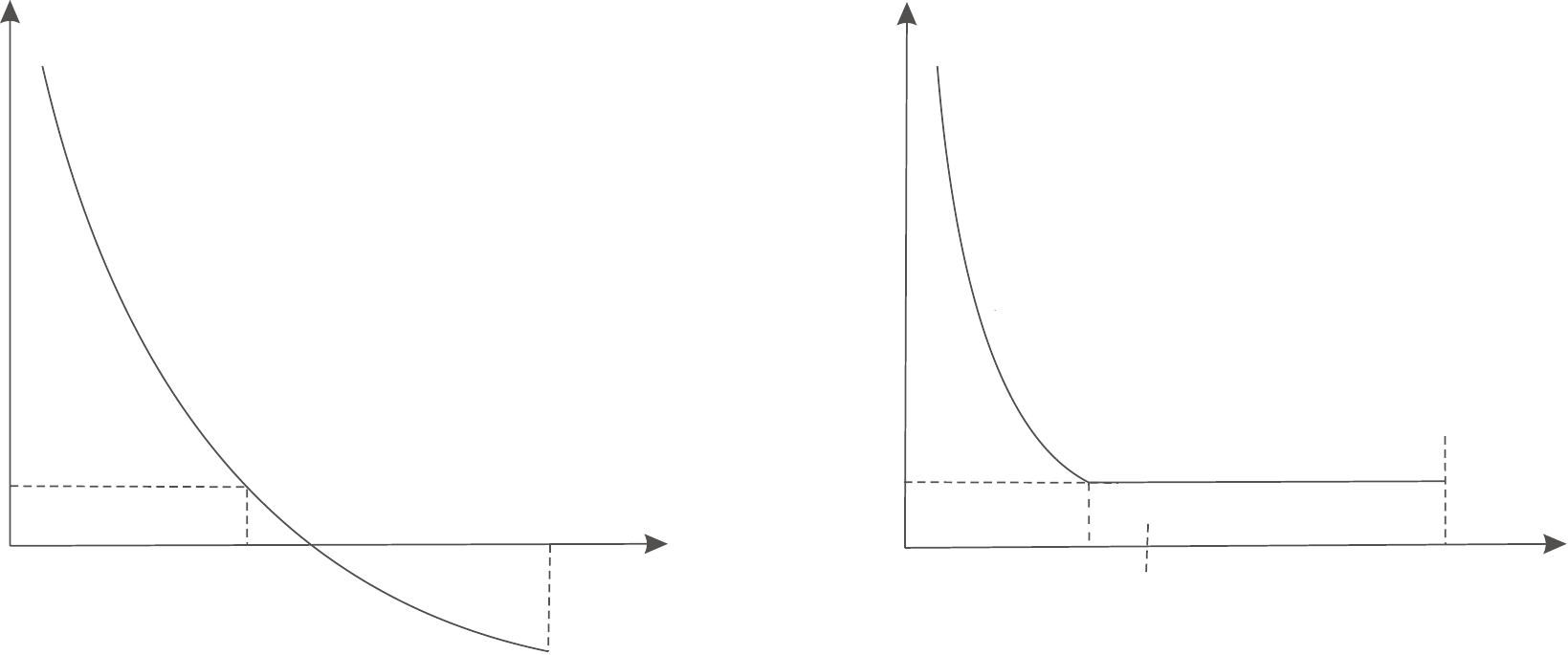}
\begin{picture}(0,0)
\put(-5,10){$\phi$}
\put(-35,10){$\pi$}
\put(-95,10){$\phi_{\rm E3}$}
\put(-118,10){$\phi_{\rm O3}$}
\put(-150,10){$0$}
\put(-345,10){$0$}
\put(-300,10){$\phi_{\rm E3}$}
\put(-200,10){$\phi$}
\put(-230,30){$\pi$}
\put(-375,38){$\frac{h(\phi_{\rm E3})}{g_s\, N}$}
\put(-370,130){$\frac{h(\phi)}{g_s\, N}$}
\put(-175,130){$\frac{h(\phi)}{g_s\, N}$}
\end{picture}
\caption{Illustration of the function $h(\phi)/g_sN$ on the $S^6$ model of the compact space. Left: The function is normalized to take a small value at the E3 position $\phi_{\rm E3}$ and consequently goes negative in a significant part of space. Right: The arrangement of all O3 planes on a 5-sphere at $\phi_{\rm O3}< \phi_{\rm E3}$ may avoid the singular-bulk problem.
\label{wf}}
\end{figure}
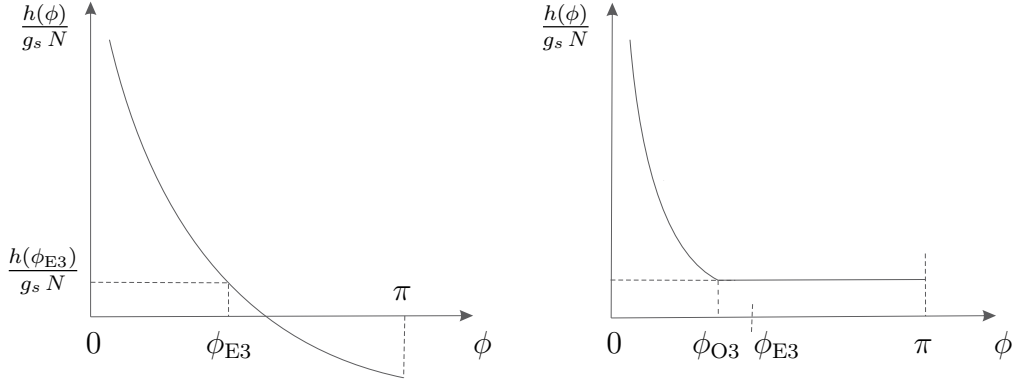

To satisfy Gauss's law, we also require negatively-charged sources corresponding to the O3 planes. Let us assume that they are equidistantly distributed across the whole 6-sphere. Moreover, if $N$ is large we may replace these $4N$ point sources by a homogeneous charge distribution. The warp factor equation (\ref{epp}) then becomes
\begin{equation}
\frac{1}{\sin^5(\phi)}\left[\sin^5(\phi)\, h(\phi)^\prime\right]^\prime = -\frac{g_sN \,\delta(\phi)}{V(S^5)\sin^5(\phi)} + \frac{g_sN}{V(S^6)}\,,
\end{equation}
with $V(S^n)$ the volume of the unit-radius $n$-sphere. This is the same as
\begin{equation}
\pi^3\left[\sin^5(\phi)\, h(\phi)^\prime\right]^\prime = -g_sN\,\left(\delta(\phi)-\frac{15}{16}\sin^5(\phi)\right)\,.
\end{equation}
The r.h.~side averages to zero on the interval $(0,\pi)$, as required by tadpole cancelation. After this equation has been integrated once, the constant of integration must be adjusted such that the r.h.~side vanishes at $\phi\to\pi$. The second integral then gives an expression for $h(\phi)$ which behaves as
\begin{equation}
h(\phi)\simeq \frac{g_sN}{4\pi^3 \phi^4}
\end{equation}
at $\phi\ll 1$ and approaches a constant value near the south pole (at $\phi\to \pi$). This is illustrated on the l.h.~side of Fig.~\ref{wf}.
As this figure suggests, it will be convenient to think of the warp factor as $h=g_sN h_0$, where $h_0$ is an ${\cal O}(1)$ function diverging at $\phi\to 0$ and approaching a constant at $\phi\to\pi$.

In principle, we are free to shift the function $h$ by an arbitrary constant (which could be identified with the volume modulus in an actual string compactification). Analogously to KKLT, we fix this freedom by demanding (cf.~\eqref{mod})
\begin{equation}
h(\phi_\text{E3}) \sim \frac{N}{M^2 V(S^4)\sin^4(\phi_{\rm E3})}\,.\label{hcond}
\end{equation}
Here we have modeled the E3 brane as a maximal-radius 4-sphere embedded in the 5-sphere at $\phi=\phi_{\rm E3}$. The unwarped E3 volume is hence $V(S^4)\sin^4(\phi_{\rm E3})$. We can then rewrite $h$ as
\begin{equation}
h(\phi) \sim h(\phi)-h(\phi_\text{E3})+\frac{N}{M^2 V(S^4)\sin^4(\phi_\text{E3})}
\end{equation}
or
\begin{equation}
h(\phi) \sim g_sN\left(h_0(\phi)-h_0(\phi_\text{E3})+\frac{1}{g_sM^2 V(S^4)\sin^4(\phi_\text{E3})}\right). \label{xyz}
\end{equation}
For generic $\phi_{\rm E3}$ and $g_sM^2\gg 1$, the third term inside the brackets is parametrically small. At the same time, the function $h_0(\phi)-h_0(\phi_\text{E3})$ is smooth and falls monotonically, approaching an ${\cal O}(1)$ negative value at $\phi\to \pi$. Hence the inequality
\begin{equation}
|h_0(\phi) - h_0(\phi_\text{E3})| \,\geq \, \frac{1}{g_sM^2 V(S^4)\sin^4(\phi_\text{E3})}
\end{equation}
holds in a significant part of our space, making $h$ negative and our geometry singular in that region.

Our argument has the drawback that the E3-brane model is too simplistic. This can not be improved as long as we model the total space by an $S^6$ since the latter has no non-trivial 4-cycle.  As a result, our toy model offers the escape route of placing the E$3$ near the south pole, such that $|\pi-\phi_{\rm E3}|\ll 1$. This helps in two ways: First, the E3 volume $V(S^4)\sin^4(\phi_{E3})$ becomes small. Second, since the E3 sits close to the minimum of the warp factor, $h_0(\phi)-h_0(\phi_\text{E3})$ cannot drop much below zero. Both effects prevent the r.h.~side of \eqref{xyz} from becoming negative and the problem disappears.
Whether these loopholes persist in the Calabi-Yau case is not obvious. There, the E3 position is fixed dynamically and its volume is bounded from below since we are wrapping a non-trivial 4-cycle. We will discuss the possibility of a small E3 volume in Sect.~\ref{sec:escapes2} and show that it comes with its own problems.
Placing the E3 close to a minimum of $h$ is difficult in the Calabi-Yau case as well.
In particular, although the E3 with the smallest action extremizes $h$ along the transverse space (for negligible worldvolume flux) \cite{Bianchi:2012kt}, $h$ generically also varies along the directions parallel to the E3. Recall further that $h$ has no minima away from points with negative D3 charge. An idea to possibly avoid these problems is discussed in Sect.~\ref{sec:curv}.

Apart from placing the E3 near the south pole, the toy model offers another way out which might also be relevant for realistic geometries (see also Sect.~\ref{sec:screen}): Let us assume that the O3 planes are {\it not} scattered over the whole $S^6$ but rather all sit on an $S^5$ at some fixed angle $\phi = \phi_\text{O3}$. Moreover, let them be distributed equidistantly on that $S^5$, such that for $N\gg 1$ their effect is similar to that of a homogeneous negative charge distribution on the $S^5$. The warp factor equation thus becomes
\begin{equation}
\pi^3 \left[\sin^5(\phi)\, h(\phi)^\prime\right]^\prime = -g_sN \delta(\phi) + g_sN \delta(\phi-\phi_\text{O3}).
\end{equation}
For $\phi<\phi_\text{O3}$, it is straightforward to calculate explicitly
\begin{equation}
h(\phi)=-\int \d\phi \,\frac{g_sN}{\pi^3 \sin^5(\phi)}\,+\,\mbox{const.}
\end{equation}
For $\phi>\phi_\text{O3}$, the function $h(\phi)$ is constant.
The reason is that, in this region, the D3 charge of the throat is completely screened by the 5-sphere of negative charge. Let us now assume that the E3 cycle is located in this region of constant $h$. Then, as illustrated on the r.h.~side of Fig.~\ref{wf}, the choice of an appropriate additive constant allows us to satisfy the KKLT criterion (\ref{hcond}) while avoiding any negative values.

In summary, we hope that our toy model makes two points: First, it illustrates explicitly how the strong variation of $h$ together with the KKLT condition generically leads to singularities. But second, it also shows that very special arrangements of O3 planes and the E3 cycle may in principle avoid the problem.
Whether these and other escape routes can be realized in concrete Calabi-Yau geometries will be discussed in the next section.

\section{Possible escape routes}
\label{sec:escapes}

In this section, we discuss ideas to avoid the singular-bulk problem. For simplicity, we assume $h^{1,1}_-=0$ throughout this section. We first analyze to which extent one may circumvent the problem in models with $h_+^{1,1}=h^{1,1}=1$. Models with $h^{1,1}\neq 1$ will be discussed further below.

\subsection{Models with $h^{1,1}=1$}

\subsubsection{Gaugino condensation with $N_C\gg 1$}

We recall that the singular-bulk problem hinges upon $g_sM^2$ being large in the regime of parametric control, $g_sM^2\gg 1$. This is the same parameter which underlies the `throat-gluing problem' of \cite{Carta:2019rhx}, see also our Sect.~\ref{sec:setup}. As can be seen from that section, replacing the E3-instanton effect by $SU(N_C)$ gaugino condensation leads to the replacement $g_sM^2\,\,\to \,\, g_sM^2/N_C$. However, as already noted in \cite{Carta:2019rhx}, this can only alleviate the problem slightly since, according to \cite{Louis:2012nb}, D7 tadpole constraints limit $N_C$ roughly by $N_C\lesssim {\cal O}(10)\,h^{1,1}$. Therefore, $N_C$ can be of the same order as $g_sM^2$ only at the boundary of control (i.e., for $g_sM^2 \simeq 12$) but not for $g_sM^2\gg 12$. So one is forced into the regime of $h^{1,1}\gg 1$ which, as we discuss below, has its own issues. A caveat is that the rule $N_C\lesssim {\cal O}(10)\,h^{1,1}$ was verified in \cite{Louis:2012nb} by studying a (very large) set of Calabi-Yau manifolds, so it might in principle be violated in examples not included in this set.

\subsubsection{D3 tadpole from 7-brane curvature}
\label{sec:curv}

While we assumed for definiteness and simplicity that the required negative contribution to the D3 tadpole comes from a large set of O3 planes, this is not the only option. An alternative is the corresponding contribution from an integral over the curvature of O7 planes or D7 branes. Our arguments about unavoidable large regions with $h(y)<0$ from Sect.~\ref{sec:basic} (or the coarse-grained $h_c(y)<0$ in Sect.~\ref{sec:sing}) still go through. It is also expected that these large negative regions are distinct from the string-theoretically resolved singularities near the 7-branes. However, a more careful study might be justified since the latter singularities are of complex co-dimension one (rather than co-dimension three in the O3 case). So maybe it would be simpler to localize a large fraction of the E3 volume near such singularities and hence at small values of $h(y)$. Moreover, one could imagine a situation where the gaugino condensate providing the non-perturbative effect occurs on the same brane stack the curvature of which dominates the D3 tadpole. This could make it easier to have the gauge-theory-brane volume be mostly in a small-$h$ region. However, it is not clear to us which effective value of $h$ (the actual value being divergent on the brane) one should use in this case. We have to leave a careful study of this problem to future research.

It is also interesting that, at least naively, curvature corrections to the D7-brane action can weaken the condition \eqref{tkm} that was crucial to derive the problematic inequality \eqref{ineq}. In particular, consider a supersymmetrically embedded D7-brane stack wrapped on a 4-cycle with Euler number $\chi(\Sigma)/24= N/N_C$. Each brane then carries a D3 charge $-N/N_C$ induced by curvature corrections to its CS action, and one can verify that it also receives $\mathcal{O}(\frac{N}{N_C})$ curvature corrections to its DBI action \cite{Giddings:2001yu}. The latter have the opposite sign as the leading-order term proportional to the 4-cycle volume, leading to a corrected gauge kinetic function $f_\text{D7}\sim T-\frac{N}{N_C}S$ (with $S$ the axio-dilaton) \cite{Haack:2006cy} and, hence, a non-perturbative superpotential $\sim \exp\left(-\frac{2\pi}{N_C}(T-\frac{N}{N_C}S)\right)$.\footnote{The same conclusion applies with $N_C=1$ if the superpotential is generated by an instanton with $\chi(\Sigma)/24=N$.} Following Sect.~\ref{sec:setup} and neglecting numerical factors, we thus find $\text{Re}(T)-\frac{N}{N_C}\text{Re}(S)\simeq \frac{N_CN}{g_sM^2}$. There are now two cases: Either the second term on the l.h.~side is much smaller than the first one or it is not. In the former case, nothing changes parametrically and the problematic condition \eqref{tkm} still holds. In the latter case, one finds that the condition is relaxed to $\text{Re}(T) \simeq N/g_sN_C$.
However, it is not clear to us whether this would correspond to a trustworthy vacuum, as the naively leading term and the curvature corrections in the DBI action would then be of the same order, which typically signals a breakdown of the $\alpha^\prime$ expansion.

Let us finally note that cancelling the tadpole using curved D7 branes and O7 planes may lead to further parametric control problems related to the requirement of large $N$, as we discuss in more detail in Sect.~\ref{sec:tadpole}.

\subsubsection{Screened KS charge}
\label{sec:screen}

In geometries where the orientifold fixed points are all located near the conifold region, the negatively charged O-planes may effectively screen the charge $N$ sitting at the bottom of the conifold (see also the discussion at the end of Sect.~\ref{sec:toymodel}). The warp factor in the bulk does then not see any monopole charge and consequently varies very slowly. This is a highly ungeneric situation, but it is not clear to us why there would be any problem in principle with such a configuration. It would therefore be interesting to explore this possibility further in explicit geometries.

\subsection{Models with $h^{1,1}>1$}
\label{sec:escapes2}

Another possible route to avoid the singular-bulk problem are models with several Kahler moduli. This has the advantage that one may be able to generate a large hierarchy between a 4-cycle that dominates the Calabi-Yau volume and the volume of another 4-cycle $\Sigma$ supporting a non-perturbative effect. In terms of volumes in $\tilde g$, this corresponds to the regime $\tilde V_\Sigma \ll 1$ (and $\mathcal{\tilde V}=1$ by definition). As can be seen from Sect.~\ref{sec:basic}, the singular-bulk problem is alleviated for small $\tilde V_{\Sigma}$ since the problematic large factor $g_sM^2$ in \eqref{ineq} is then replaced by $g_sM^2 \to g_sM^2 \tilde V_{\Sigma}$. The singular-bulk problem therefore disappears if
\begin{equation}
\tilde V_{\Sigma} \lesssim \frac{1}{g_sM^2} \ll 1. \label{ratio}
\end{equation}
In such a regime, the bulk is only weakly warped (i.e., $|\tilde{\partial h}|/h \lesssim 1$) and no dangerous singularities are expected to arise.
For later convenience, let us also write the condition in terms of the Einstein-frame 4-cycle volume $\tau_\Sigma$ and the Einstein-frame Calabi-Yau volume $\mathcal{V}$ in the physical metric $g$. Using that $\tilde V_{\Sigma}=\tilde V_{\Sigma}/\mathcal{\tilde V}^{2/3}\approx \tau_\Sigma/\mathcal{V}^{2/3}$ at weak warping, we find
\begin{equation}
\frac{\tau_\Sigma}{\mathcal{V}^{2/3}} \lesssim \frac{1}{g_sM^2} \ll 1. \label{ratio2}
\end{equation}
As we will see below, it is difficult to satisfy \eqref{ratio2} if the Kahler moduli are stabilized using only non-perturbative effects. However, we find that the singular-bulk problem is relaxed if also $\alpha^\prime$ corrections are taken into account as in the LVS. To see this, we now go through several scenarios in detail:

\subsubsection{Standard KKLT}
\label{sec:kklt}

The simplest possibility is to consider the case where all Kahler moduli are stabilized by $F$-term conditions as in the standard KKLT scenario (before uplifting to dS using an $\overline{\text{D3}}$ brane as usual). We thus have to satisfy $D_i W = 0$ for all $i=1,\ldots,h^{1,1}$, where we assume a non-perturbative term $\sim A_i e^{-a_i\tau_i}$ in the superpotential for each modulus.
Here and in the following, we denote the Einstein-frame 4-cycle volumes (in $g$) by $\tau_i$.
It is straightforward to see that, assuming $A_i \sim\mathcal{O}(1)$, the $F$-terms then imply that all 4-cycle volumes are of the same order, i.e., $a_1\tau_1 \sim a_2\tau_2 \sim \ldots $ up to log corrections. It is therefore not possible to generate a large hierarchy between a 4-cycle volume and $\mathcal{V}^{2/3}$.
A way around this conclusion may be large $h^{1,1}\gg 1$, possibly together with large ratios $a_i/a_j$ (i.e., large gauge groups). We will discuss this possibility further below in Sects.~\ref{sec:largeh11} and \ref{sec:largeh11nc}. In all other cases, we conclude that \eqref{ratio2} cannot be satisfied.

\subsubsection{Stabilization with a single non-perturbative effect}
\label{sec:bobkov}

Alternatively, we can consider a stabilization scheme where one non-perturbative effect stabilizes all Kahler moduli as proposed in \cite{Bobkov:2010rf}. The idea is to have a superpotential of the form $W=W_0 + Ae^{-a_\Sigma T_\Sigma}$, where $T_\Sigma= n^iT_i$ with $n^i\in \mathbb{N}^+$ is some linear combination of the Kahler moduli.
The $F$-term conditions are then $D_\Sigma W = 0$ and $K_{a} = 0$, where we denote by $T_a$ with $a=1,\ldots,h^{1,1}-1$ the moduli combinations orthogonal to $T_\Sigma$. It was shown in \cite{Bobkov:2010rf} that $K_a = 0$ implies $t^i \sim n^i$ for the 2-cycle volumes $t^i$ dual to the 4-cycle volumes $\tau_i$. For appropriately chosen $n^i$, one can therefore generate a hierarchy between the $t^i$. Because of $\tau_i = \frac{1}{2}\kappa_{ijk}t^jt^k$, this may also yield a hierarchy between the 4-cycle volumes. However, it turns out that this does not alleviate the singular-bulk problem in the present case. The basic issue is that the largest 4-cycle always appears in the exponent of the non-perturbative term because $T_\Sigma$ includes a positive sum over all 4-cycle volumes. Since the E3 wraps a positive combination of all 4-cycles including the largest one, it is clear that the E3 volume must be of the same order as $\mathcal{V}^{2/3}$. We therefore conclude that \eqref{ratio2} cannot be satisfied.

\subsubsection{$D$-term stabilization}

Another possibility is to admit a stabilization of the Kahler moduli by a combination of $F$-terms and $D$-terms.
The general $D$-term potential is (see, e.g., \cite{Villadoro:2005yq, Jockers:2005zy, Haack:2006cy})
\begin{equation}
V_D \sim D_a D^a, \qquad D_a = q_{ia}\frac{D_i W}{W}, \qquad q_{ia} = f^j\kappa_{jia}, \label{dterm0}
\end{equation}
where the $f^j$ are worldvolume fluxes on a D7 brane wrapped on some 4-cycle $S_a$. For simplicity, we assume here that the vevs of charged matter fields are zero (see, e.g., \cite{Jockers:2005zy, Haack:2006cy, Blumenhagen:2007sm} for a discussion of such terms).
Note that each $D$-term is proportional to a linear combination of the $F$-terms. Stabilizing $n< h^{1,1}$ Kahler moduli by $D$-term conditions and the remaining $h^{1,1}-n$ Kahler moduli by $F$-term conditions therefore implies $D_{i}W=0$ for all $i$.

A consistent $D$-term furthermore requires \cite{Villadoro:2005yq}
\begin{equation}
q_{ia}\frac{\partial_i W}{W} \in \mathbb{R} \quad \forall a
\end{equation} off-shell for all field values. Because of $W_0 \neq 0$ (and higher instanton corrections), this is not compatible with a non-perturbative term $\sim e^{-T_i}$ in $W$ unless $q_{ia}\partial_i W=0 \,\,\, \forall a$.
An alternative argument leading to the same conclusion was given in \cite{Blumenhagen:2007sm}.
The $D$-terms thus reduce to $n$ conditions of the form
\begin{equation}
q_{ia}K_i =0. \label{dterm}
\end{equation}
We thus arrive at a situation very similar to the one discussed in Sect.~\ref{sec:bobkov}. In particular, only $h^{1,1}-n$ linear combinations of the $T_i$ appear in $W$ while the remaining $n$ moduli are fixed by constraints involving only $K$.

To see why this does not resolve the singular-bulk problem, let us choose the 2-cycle volumes $t^i$ such that they are a basis of the Kahler cone with $t^i>0$. Because of $D_i W = 0$ and $K_i W \sim t^i$,\footnote{Here, we used $W\neq 0$ and the relation $K_i = - t^i/2\mathcal{V}$, which follows from $K = -2\ln \mathcal{V}+\ldots$ and the fact that $\mathcal{V}$ is a homogeneous function of the $\tau_i$ of degree $3/2$ (see, e.g., \cite{Bobkov:2010rf}). It would be interesting to analyze whether the crucial conclusion $K_iW \sim t^i \neq 0$ can be avoided for non-zero matter vevs.} we then have $\partial_i W \neq 0$ for all $i$. Hence, the exponents of the instanton terms in $W$ are linear combinations involving \emph{every} $\tau_i$. For instantons contributing to $W$, the coefficients of these combinations are positive integers (up to an overall factor $-2\pi$). The argument is now completely analogous to our discussion at the end of Sect.~\ref{sec:bobkov}. In particular, whatever hierarchy between the $\tau_i$ we generate through the $D$-term constraints \eqref{dterm} will not help because the largest basis 4-cycle necessarily appears in at least one of the exponents in $W$. There must therefore be an instanton that wraps the largest 4-cycle
and we find again that \eqref{ratio2} cannot be satisfied.\footnote{One may more generally attempt to stabilize some of the moduli using $D$-terms and the remaining ones non-supersymmetrically, i.e., impose $q_{ia}K_i=q_{ia}\partial_i W=0$ for some linear combinations and allow $D_i W \neq 0$ for the others. We refrain from a general analysis here but note that this idea is not successful for $h^{1,1}=2$. The reason is again that the larger 4-cycle necessarily appears in the exponent of the non-perturbative term in $W$.}

\subsubsection{Parametrically large $h^{1,1}$}
\label{sec:largeh11}

Yet another possibility is to consider compactifications with $h^{1,1}\gg 1$. To see why this might be promising, let us again assume that the Kahler moduli are stabilized as in one of the KKLT variants discussed above. We have seen that one of the exponents in the non-perturbative superpotential then contains the largest basis 4-cycle. We therefore have $\tau_i \lesssim \tau_\Sigma$ for all $i$ and for some $\Sigma$ supporting one of the non-perturbative effects.
Analogous arguments to those of Sect.~\ref{sec:setup} then imply (see also \eqref{ret})
\begin{equation}
\tau_\Sigma \sim \frac{N}{g_sM^2}. \label{taus}
\end{equation}
We also use a result of \cite{Demirtas:2018akl}, which states that the volume of the largest basis 4-cycle $\tau_\text{last}\equiv \text{max}(\tau_i)$ and the Calabi-Yau volume itself must scale at least like
\begin{align}
\tau_\text{last} &\sim(h^{1,1})^p && 3.2 \lesssim p \lesssim 4.3, \notag \\
\mathcal{V} &\sim(h^{1,1})^q && 6.2 \lesssim q \lesssim 7.2 \label{scalings}
\end{align}
in regions of the Kahler cone where string corrections are under control. This was shown in \cite{Demirtas:2018akl} for a large set of Calabi-Yau manifolds with $h^{1,1}\gg 1$.
For concreteness, let us assume that $p$ and $q$ take values in the middle of the provided ranges. This seems plausible since \cite{Demirtas:2018akl} derived the bounds on $p$ and $q$ using two cones that either contain or are contained in the actual Kahler cone. We thus find $q-\frac{3}{2}p\approx 1$ and therefore
\begin{equation}
\frac{\tau_\Sigma}{\mathcal{V}^{2/3}}\gtrsim (h^{1,1})^{p-2q/3} \sim (h^{1,1})^{-2/3}. \label{ratio3}
\end{equation}
An alternative estimate coming to the same conclusion is as follows. Let us assume that the Calabi-Yau has $\mathcal{O}(h^{1,1})$ non-zero triple-intersection numbers which take $\mathcal{O}(1)$ values, in agreement with another result of \cite{Demirtas:2018akl}. Using this together with $\mathcal{V} = \frac{1}{6}\kappa_{ijk}t^it^jt^k$ and $\tau_i= \frac{1}{2}\kappa_{ijk}t^jt^k$, we find $\mathcal{V} \lesssim h^{1,1}\tau_\Sigma^{3/2}$ at large $h^{1,1}$, reproducing \eqref{ratio3}.

We thus see that we can achieve the desired hierarchy \eqref{ratio2} if $h^{1,1}\gtrsim (g_sM^2)^{3/2}$. Recall that a controlled uplift in KKLT requires $g_sM\gtrsim 1$ and $M\gtrsim 12$. It therefore seems that we can resolve the singular-bulk problem on manifolds with $h^{1,1}\gtrsim 12^{3/2} \approx 42$.

However, this idea is likely incompatible with tadpole cancelation. Indeed, using \eqref{scalings} in \eqref{taus} yields
\begin{equation}
\frac{N}{g_sM^2} \gtrsim (h^{1,1})^{3.2} \gtrsim (g_sM^2)^{4.8}.\label{die}
\end{equation}
Therefore, $N \gtrsim (g_sM^2)^{5.8} \gtrsim 12^{5.8} \approx 1.8\cdot 10^6$, which is way too large even at the boundary of control.\footnote{For simplicity, we have not kept track of the (rather small) numerical prefactors that appear in the scaling laws of \cite{Demirtas:2018akl} for $\tau_\text{last}$ and $\mathcal{V}$. One can verify that taking into account these factors in \eqref{ratio3} and \eqref{die} would make our bound on $N$ even stronger.} As stated before, the largest known tadpole including F-theory models is $7.5\cdot 10^4$ \cite{Taylor:2015xtz}.

We stress that this argument is not a proof that large $h^{1,1}$ cannot work. Indeed, the bound on $N$ is rather sensitive to the precise exponents assumed in the scalings \eqref{scalings}.
In particular, significantly smaller values for $N$ become possible if the true $q$-value is close to its upper bound and the true $p$-value is close to its lower bound. One should also keep in mind that the results of \cite{Demirtas:2018akl} are ``experimental'', i.e., they were found to hold in many examples but could in principle be violated in manifolds not included in the set. Nevertheless, our calculation demonstrates how difficult it is to solve the singular-bulk problem without creating another control problem somewhere else.

\subsubsection{Combining parametrically large $h^{1,1}$ and $N_C$}
\label{sec:largeh11nc}

Let us now try to combine the two potential escape routes of large $N_C$ and large $h^{1,1}$. As explained earlier, one may write $N_C = \beta h^{1,1}$ with $\beta$ restricted by $\beta \lesssim {\cal O}(10)$ \cite{Louis:2012nb}. We recall from Sect.~\ref{sec:basic} that the factor $N_C$ may be introduced by systematically replacing $g_s M^2$ with $g_s M^2/N_C$. Thus, (\ref{die}) turns into
\begin{equation}
\frac{N\,\beta h^{1,1}}{g_sM^2} \gtrsim (h^{1,1})^{3.2} \gtrsim \left(\frac{g_sM^2}{\beta h^{1,1}}\right)^{4.8}.
\end{equation}
This implies $h^{1,1}\gtrsim(g_sM^2/\beta)^{3/5}$ and hence $N\gtrsim (h^{1,1})^{2.2}g_sM^2/\beta\gtrsim 
(g_sM^2/\beta)^{2.3}$. Even taking the conservative value $\beta\simeq 1$, the resulting tadpole is not prohibitively large.
Instead, any $g_sM^2 \lesssim N^{0.43}\beta$ is compatible with the constraints. Depending on the values for $N$ and $\beta$ in a given model, this may allow reasonably good control, i.e., $g_sM^2 \gg 12$.
Thus, while relatively involved, this appears to be a realistic escape route.

However, note that having only a single brane stack with large $N_C$ would not be sufficient for this to work. Indeed, as explained in Sect.~\ref{sec:kklt}, moduli stabilization according to KKLT yields $a_1\tau_1 \sim a_2\tau_2 \sim \ldots \sim N/g_sM^2$ for all $\tau_i$. All 4-cycles without large stacks (i.e., with $a_i \sim \mathcal{O}(1)$) therefore satisfy $\tau_i \sim N/g_sM^2$. It was furthermore pointed out in \cite{Demirtas:2018akl} that not only the largest but in fact \emph{most} basis 4-cycles need to scale non-trivially with $h^{1,1}$ in order to maintain perturbative control.\footnote{However, the precise scaling was stated in \cite{Demirtas:2018akl} only for the largest 4-cycle volume $\tau_\text{last}$, cf.~\eqref{scalings}.} We therefore expect that our arguments in Sect.~\ref{sec:largeh11} can be applied to almost any 4-cycle without a large brane stack. This means that the singular-bulk problem persists unless an inequality similar to \eqref{die} holds, which may lead to unacceptably large tadpoles as described below that equation.
The escape route of combining large $h^{1,1}$ and large $N_C$ can therefore only work if most of the basis 4-cycles support large brane stacks of the order $N_C\sim \beta h^{1,1}$. This yields a total gauge-group rank $\text{rk}(SU(N_C)^{\mathcal{O}(h^{1,1})})\sim \beta (h^{1,1})^2$. It is not clear to us whether there are manifolds on which such large gauge groups are compatible with D7 tadpole cancelation.\footnote{Note that the bounds found in \cite{Louis:2012nb} apply to the maximal gauge-group rank on a single stack, not to the total gauge-group rank.}

Alternatively, we may reconsider the scenario of \cite{Bobkov:2010rf} discussed in Sect.~\ref{sec:bobkov}, where a single non-perturbative effect stabilizes all Kahler moduli. However, as we will now show, the combination of large $h^{1,1}$ and large $N_C$ is then not sufficient to solve the singular-bulk problem. As explained before, the scenario implies that $\tau_\Sigma$ is a linear combination involving all basis 4-cycles, $\tau_\Sigma \sim n^i\tau_i$ with $n^i\in \mathbb{N}^+$. Assuming that a brane stack with large $N_C$ is wrapped on $\Sigma$, we find
\begin{equation}
n^i\tau_i \sim \frac{NN_C}{g_sM^2} \sim \frac{N\beta h^{1,1}}{g_sM^2}. \label{die2}
\end{equation}
Since the left-hand side involves a sum over $h^{1,1}$ cycle volumes, it is clear that most $\tau_i$ must satisfy $\tau_i \lesssim N\beta/g_sM^2$ or else \eqref{die2} would be violated. As stated above, many of these cycle volumes scale non-trivially with $h^{1,1}$ in a controlled regime. Following the arguments of Sect.~\ref{sec:largeh11}, we are thus led back to an inequality like \eqref{die} (up to a factor $\beta$) and the problems described there.

We conclude that, even allowing both $h^{1,1}$ and $N_C$ to be large, it is surprisingly difficult to construct a viable model in which the singular-bulk problem is avoided.

\subsubsection{Large-volume scenario (LVS)}

Let us finally discuss LVS moduli stabilization \cite{Balasubramanian:2005zx} (for recent work and more references see e.g.~\cite{AbdusSalam:2020ywo}). For the standard example of a swiss-cheese manifold with 2 Kahler moduli $\tau_1$ and $\tau_2$,
an LVS minimum exists for
\begin{equation}
\mathcal{V} \sim \tau_2^{3/2} \sim e^{a_1\tau_1}|W_0| \sqrt{\tau_1}, \qquad \tau_1 \sim \frac{\xi^{2/3}}{g_s}, \label{v}
\end{equation}
where $\xi \approx -2.4\cdot 10^{-3} \chi(X)$ and $\chi(X)$ is the Euler characteristic of the Calabi-Yau 3-fold.
Assuming a dS uplift by an $\overline{\text{D3}}$ brane, the uplift term needs to be of the same order as the terms in the AdS potential, as in the KKLT scenario (cf.~Sect.~\ref{sec:setup}). One can check that this yields the same condition $\tau_1 \sim N/g_sM^2$ that we had before (up to $\mathcal{O}(1)$ factors and log corrections). Since $\tau_1 \sim \xi^{2/3}/g_s$, we furthermore require $N\sim M^2\xi^{2/3}$.
However, we also see from \eqref{v} that the ratio \eqref{ratio2} (with $\tau_\Sigma=\tau_1$) is now exponentially small. We therefore expect no singular-bulk problem in the LVS.

\section{Further control issues}
\label{sec:tadpole}

In the previous sections, we argued that the requirement $g_sM^2\gg 1$ implies serious control issues for flux compactifications admitting a KKLT-like dS uplift. The purpose of the present section is to analyze a different type of problem, which is related to the requirement of a large tadpole $N$.

Let us first recall how the requirement $N\gg 1$ arises. This follows because, according to \eqref{mod}, the string-frame volume of the 4-cycle wrapped by the instanton is of the order $N/M^2$. For perturbative control, this volume needs to be large in string units and therefore
\begin{equation}
N \gg M^2 \gg 1. \label{tadpole}
\end{equation}
We also recall that the Einstein-frame 4-cycle volume satisfies
\begin{equation}
\tau_\Sigma \sim \frac{N}{g_sM^2} \ll N. \label{tau}
\end{equation}

The point we now want to make is that \eqref{tadpole} implies a complicated topology of the Calabi-Yau manifold, which in turn may lead to control issues. Intuitively, the problem is that the limit of large $N$ does not correspond to the usual large-volume limit in which one expects a parametrically good control over $\alpha^\prime$ corrections. Instead, we will see that taking $N$ large means taking the volume large while at the same time increasing the number of cycles on the manifold. The danger is now that the number of cycles grows so fast at large $N$ that some of their volumes shrink below unity even though the volume of $\Sigma$ and the total Calabi-Yau volume become large.

Here, we have to warn the reader that our arguments in this section are not fully conclusive and more work is needed to explore these ideas further. Let us in the following nevertheless present some evidence that the large-$N$ limit may indeed be problematic.

To illustrate our claim, we focus on a simple class of F-theory models where the tadpole is generated by a curved D7 brane and O7 plane such that we have a description in terms of a smooth elliptically fibered Calabi-Yau 4-fold $Y$. We will also restrict to the weak-coupling limit such that the base is a quotient of a Calabi-Yau 3-fold and $g_s$ is small and approximately constant except near the 7-branes. The tadpole is given by $N=\chi(Y)/24$, where $\chi(Y)$ denotes the Euler number of the 4-fold. It can be expressed in terms of the Hodge numbers on $Y$ as \cite{Klemm:1996ts}
\begin{equation}
\chi(Y) = 6(8+h^{1,1}(Y)+h^{3,1}(Y)-h^{2,1}(Y)).
\end{equation}
Note that $\chi(Y)$ is positive because $\chi(Y)/24 = \int_Y G_4 \w G_4 + N_\text{D3} = \int_Y \star G_4 \w G_4  + N_\text{D3} > 0$, where we used the tadpole condition together with the self-duality of the 4-form flux $G_4$. Since $h^{2,1}(Y)$ contributes negatively to $\chi(Y)$, we conclude that $N=\chi(Y)/24$ implies
\begin{equation}
h^{1,1}(Y) + h^{3,1}(Y) \gtrsim 4N. \label{hy}
\end{equation}

We now want to relate the Hodge numbers on $Y$ to the Hodge numbers on the 3-fold $X$. This works as follows (see, e.g., \cite{Denef:2008wq, Collinucci:2008pf}):
$h^{1,1}(Y)=h^{1,1}_+(X)+1$ counts the Kahler moduli, $h^{2,1}(Y)=h^{1,1}_-(X)$ counts the $B_2$ and $C_2$ axions, and $h^{3,1}(Y)=h_-^{2,1}(X)+\hat h_-^{2,0}(S)+1$ counts the complex-structure moduli, the D7-brane deformations and the axio-dilaton. Here, we denote by $S$ the 4-manifold wrapped by the D7 brane. The hat on $\hat h_-^{2,0}(S)$ indicates that this surface is generically singular in F-theory so that special care is required when computing its topological invariants \cite{Collinucci:2008pf, Braun:2008ua}. Using these relations in \eqref{hy}, we find
\begin{equation}
h^{1,1}_+(X) + h_-^{2,1}(X)+\hat h_-^{2,0}(S) \gtrsim 4N. \label{hx}
\end{equation}
We thus see that at least one of these three Hodge numbers must be of the order $N$ or larger.

Before we proceed, a comment is in order. As stated above, we have restricted our discussion to smooth 4-folds. In general singular 4-folds, the formulae relating the Hodge numbers on $Y$ and $X$ include a further term involving the rank of the gauge group.\footnote{The gauge group also involves an Abelian sector. Its rank is determined by the rank of the associated Mordell-Weil group of rational sections in elliptic Calabi-Yau manifolds, which is $\mathcal{O}(1)$ in typical examples (for a review, see \cite{Cvetic:2018bni,Weigand:2018rez}).}
However, recall that \cite{Louis:2012nb} argued that
the maximal rank of a brane stack is $N_C \sim \mathcal{O}(10)\, h^{1,1}(X)$. It is not clear to us whether the \emph{total} gauge-group rank needs to satisfy a similar bound but if it does, one cannot parametrically escape the conclusion \eqref{hx} in such models.\footnote{See also \cite{Taylor:2015xtz, Wang:2020gmi} for an explicit example where the gauge-group rank is very large and indeed of the same order as $h^{1,1}(X)$.}
Another interesting case not studied here are models with O3 singularities, which give an extra contribution to the tadpole \cite{Collinucci:2008zs}. It would be interesting to revisit the $N$ scaling of the Hodge numbers in such models.

Let us now return to the inequality \eqref{hx}. As stated before, we claim that it indicates a loss of control. This can be made most explicit if we satisfy \eqref{hx} by having $h_+^{1,1}(X) \gtrsim N$. In that case, we can use again the results of \cite{Demirtas:2018akl}.
The Calabi-Yau 3-fold then has $N$ basis 4-cycles with volumes $\tau_i$. Stabilizing them as in one of the KKLT variants discussed in Sect.~\ref{sec:escapes}, we have $\tau_i \lesssim \tau_\Sigma$ and, because of \eqref{tau}, we furthermore have $\tau_\Sigma \ll N$.\footnote{In the presence of a non-trivial gauge group with $N_C\neq 1$, we would have the weaker bound $\tau_\Sigma \ll N N_C \lesssim \mathcal{O}(10) N h^{1,1}$. This is still parametrically too small to avoid the following argument.}
On the other hand, according to \eqref{scalings}, we would require $\tau_\Sigma \gtrsim (h^{1,1}_+)^{3.2} \sim N^{3.2}$ in order to maintain perturbative control. This is clearly violated at large $N$, which confirms our claim that we lose control in spite of naively being in the large-volume regime.

We do not know whether the remaining two cases $h_-^{2,1}(X) \gtrsim N$ and $\hat h_-^{2,0}(S) \gtrsim N$ are problematic in a similar fashion.
In the first case, one may wonder whether Calabi-Yau manifolds and their 3-cycle volumes satisfy similar scaling laws at large $h_-^{2,1}(X)$ as those found in \cite{Demirtas:2018akl} at large $h_+^{1,1}(X)$. However, we are not aware of such a result in the literature. A technical complication in such an analysis could be that computing 3-cycle volumes is more difficult than 4-cycle volumes because a calibration form for 3-cycles is only known when they are special Lagrangian.

The third possibility to satisfy \eqref{hx} is to have $\hat h_-^{2,0}(S)\gtrsim N$, i.e., $\mathcal{O}(N)$ D7-brane moduli.
Naively, the simplest way to achieve this would be to consider $N$ branes instead of one, either as a stack with $N_C=N$ or distributed on $N$ homologically different 4-cycles. However, as explained before, this implies a large $h^{1,1}_+(X)\sim N$, which we already dismissed. Let us therefore stick to the case of one brane. We are thus led to models in which a single submanifold has $N$ non-trivial 2-cycles.

Assuming the D7 brane wraps a combination of basis 4-cycles with $\mathcal{O}(1)$ coefficients, its volume cannot be parametrically larger than $\tau_\Sigma$.
The D7-brane volume then satisfies $\tau_S \sim \tau_\Sigma \ll N$ because of \eqref{tau}. At first sight, this seems to be problematic: A naive estimate is that the $N$ 2-cycles will generically have sub-stringy volumes $\sim \sqrt{\tau_S/N} \ll 1$ in order to fit into the D7 brane.
However, it is not clear to us whether this is really a problem:
First, it is known that compact, orientable 4-manifolds have a property called 2-systolic freedom, i.e., they admit metrics in which volumes of non-trivial 2-cycles are \emph{not} bounded by the volume of the manifold \cite{katz}. If such metrics are allowed by the dynamics, the above ``fitting'' estimate would be wrong. Second, we do not know whether small 2-cycles on the brane would really imply a loss of control. Indeed, the 2-cycles in $H^{2,0}_-(S)$ are only non-trivial on the D7 brane but not on the ambient Calabi-Yau such that no light strings or branes could wrap on it.

Another indication of a loss of control are large curvature corrections localized on the brane worldvolume. Such corrections arise both in the cases $h_-^{2,1}(X) \gtrsim N$ and $\hat h_-^{2,0}(S) \gtrsim N$, as we will now explain.

Indeed, an alternative perspective on the topology issues discussed in this section is to consider the type IIB expression for the tadpole $N$.
In the smooth models considered here, $N$ is generated by the curvature on a single O7 and a single D7. This implies that either of them must wrap a 4-cycle with Euler number $\sim N$. The same can be concluded in more general models assuming it is not possible to distribute the tadpole among $\mathcal{O}(N)$ D7 branes/O7 planes (since according to \cite{Louis:2012nb} this would require $h^{1,1}(X)\sim N$, which we excluded above).
Considering one of the KKLT variants of Sect.~\ref{sec:escapes}, there must be a non-perturbative effect on the cycle with the large Euler number.
Then, as we already discussed in Sect.~\ref{sec:curv}, the $\alpha^\prime$ expansion on instantons or D7 branes/O7 planes wrapping that cycle can break down. We saw that the condition \eqref{tau} (or its $N_C\neq 1$ generalization $\tau_\Sigma \sim NN_C/g_sM^2$) is then naively relaxed to $\tau_\Sigma \sim N/g_sN_C$, but this is just because the curvature corrections grow to the same size as the leading volume term, suggesting that the brane action is not controlled anymore. It would be interesting to understand in more detail whether the KKLT minimum could survive this. Although the D7-brane gauge kinetic functions (and therefore the non-perturbative terms in the superpotential) are protected by non-renormalization theorems \cite{Burgess:2005jx, GarciadelMoral:2017vnz}, the same does not apply to non-BPS instantons.

We hope to come back to some of these questions in future work.

\section{Conclusions}

\label{sec:concl}

In this paper, we argued that flux compactifications of the type required in the first step of the KKLT construction of dS vacua suffer from a ``singular-bulk problem'', implying severe control issues for these vacua. Specifically, we found that in regimes admitting a controlled dS uplift (i.e., for $g_sM^2\gg 1$) a singularity develops where the warp factor becomes negative over large regions of the original Calabi-Yau manifold. We argued that this singularity is pathological and cannot be identified with the usual divergences near O-planes that are believed to be harmless in string theory. If correct, our results pose a threat to one of the leading candidates for the construction of meta-stable dS vacua in string theory.
From the 4d point of view, we expect that warping and $\alpha^\prime$ corrections blow up and thus invalidate the EFT on which the KKLT scenario is based. While we cannot exclude that the 4d EFT remains valid beyond the regime where the underlying 10d supergravity is trustworthy, this would be quite miraculous and we are not aware of any string-theory argument suggesting this.

Given the importance of the issue at stake, it will be crucial to further study the possible escape routes discussed in Sect.~\ref{sec:escapes}. Although we found that all of them come with their own problems, we cannot exclude the possibility that viable models exist for some variant of the original KKLT scenario. In particular, we were not able to rule out parametrically models with $h^{1,1}\gg 1$ in which, at the same time, each of these many Kahler moduli is stabilized by its own gaugino condensate with a large-$N_C$ gauge group. Whether models with such large gauge groups are consistent with D7 tadpole constraints remains to be seen.

Another interesting route for future work would be to investigate how the requirement of 10d consistency constrains other moduli-stabilization scenarios such as the large-volume scenario (LVS). In our present understanding, the LVS construction of dS vacua avoids the singular-bulk problem. Still, if it turned out that the LVS is the only known explicit route to stringy dS vacua, one would have to scrutinize it even more carefully. At this point, the discovery of some further problem within LVS could imply that all of string phenomenology is at stake.\footnote{The path towards a realistic string phenomenology relying on quintessence instead of de Sitter has serious issues of its own, at least in the better-understood part of the landscape \cite{Hebecker:2019csg}.}

It would also be interesting to explore the connection of our results to the (refined) dS conjecture of \cite{Ooguri:2018wrx, Obied:2018sgi, Garg:2018reu}. Indeed, our inequalities \eqref{ineq} and \eqref{ineq2} are suggestive of such a connection: If the warp factor $h$ is interpreted as a brane potential, then \eqref{ineq} and \eqref{ineq2} are formally identical to the inequalities of the conjecture. We hope to come back to this observation in the future.

Finally, we pointed out a possible problem related to the requirement of a large tadpole $N$. Our analysis in a simple class of F-theory models revealed that large $N$ implies a complicated topology of the Calabi-Yau 3-fold or of the submanifold wrapped by a D7 brane. In particular, some of the Hodge numbers are $\mathcal{O}(N)$ in this regime. We showed that this can lead to sub-stringy cycle volumes and therefore uncontrolled string corrections, even though the instanton volume and the Calabi-Yau volume are large at large $N$. While our results were not fully conclusive, it should be worthwhile to scrutinize them further, as they might pose an additional threat to dS constructions in string theory.

\section*{Acknowledgments}

We would like to thank Luca Martucci, Jakob Moritz, Pablo Soler, Alexander Westphal and Fengjun Xu for helpful discussions and correspondence. A.H. acknowledges a very helpful initial collaboration with Pablo Soler on the subject of this paper as well as earlier work with Andreas Braun, Roberto Valandro and Sven Krippendorf on Kahler moduli stabilization by D-terms. This work is supported by the Deutsche Forschungsgemeinschaft (DFG, German Research Foundation) under Germany's Excellence Strategy EXC 2181/1 - 390900948 (the Heidelberg STRUCTURES Cluster of Excellence).

\appendix

\section{Comment on the numerical prefactors}\label{nums}

Parametric estimates of geometric quantities in high dimensions tend to be affected by large powers of $(2\pi)$.
This limits the applicability to practical questions, where some of the `large' quantities are really not that large (cf.~our $M\gtrsim 12$). So it may be worthwhile being more careful with the numerical prefactors.

In particular, the reader may have been worried that a factor $(2\pi)^3$ in (\ref{rth}) was later on dismissed. Thus, let us keep that factor. At the same time, we then have to be more careful about what we call a typical Calabi-Yau radius. For example, we may model the Calabi-Yau as a torus with string-frame volume $(2\pi R_\text{CY})^6$. For the 4-cycle relevant in our context, one then has $(2\pi R_\text{CY})^4 \simeq g_s \mathrm{Re}(T) (2\pi\sqrt{\alpha^\prime})^4$.

Now let us think of gluing an $S^5$ throat, as it underlies our definition of $R_\text{throat}$ in (\ref{rth}), into that torus. Most naively, the torus is a 6d hypercube with side length $2\pi R_\text{CY}$ and opposite sides identified. It fits an $S^5$ the maximal radius of which is determined by $2R_\text{throat}=2\pi R_\text{CY}$. Thus, we have $R_\text{throat}\lesssim \pi R_\text{CY}$ and rewriting this in analogy to (\ref{gsn}) and (\ref{mmm}) gives
\begin{equation}
\frac{8 g_s N}{\pi^3}\lesssim \frac{2 N_C K}{3M}
\end{equation}
or
\begin{equation}
\frac{12}{\pi^3} \lesssim \frac{1}{g_s M^2}\lesssim \frac{1}{M}\,.
\end{equation}
We see that the various numerical prefactors do not conspire to upset the naive estimate: the l.h.~side is really an ${\cal O}(1)$ number.

Alternatively, we may try to model the Calabi-Yau which should fit the throat by a 6-sphere instead of a torus. Thus, we define $(16/15)\pi^3R_\text{CY}^6\simeq (g_s\mathrm{Re}(T))^{3/2} (2\pi\sqrt{\alpha^\prime})^6$. The maximal size of the $S^5$ is now simply determined by $R_\text{throat}\lesssim R_\text{CY}$. As a result, \eqref{mmm} then turns into
\begin{equation}
\left(\frac{16}{15}\right)^{2/3}\frac{3}{4\pi } \lesssim \frac{1}{g_s M^2}\lesssim \frac{1}{M}\,.\end{equation}
The problem is still there but, since our presumed ${\cal O}(1)$ number on the l.h. side is only about 1/4, it is less pronounced.

Let us also be more precise about the condition $g_sM\gtrsim 1$ used in the second inequality above. It comes from the radius of the $S^3$ at the bottom of the throat, which is determined by $R(S^3)^2/\alpha' \simeq 0.93\, g_s M$ \cite{Kachru:2002gs, Herzog:2001xk}. Thus, the value $g_s M = 1$ happens to be very close to the special situation where a maximal circle in the $S^3$ has T-self-dual radius. Of course, we do not know whether this is really the point where the supergravity approximation breaks down. As an alternative estimate of how small a compact space is allowed to become, let us consider the famous BBHL $\alpha'^3$ correction \cite{Becker:2002nn} to the Kahler potential, $-2\ln({\cal V})\to-2\ln({\cal V}+\xi)$.
The parameter $\xi$ is determined by the Euler number $\chi(X)$ of the Calabi Yau 3-fold and $\mathcal{V}$ denotes the Calabi-Yau volume in units of $2\pi\sqrt{\alpha^\prime}$.\footnote{
Concretely,
$\xi=-\chi(X)\zeta(3)/4(2\pi)^3$ if the volume is measured in the string frame.
}
To obtain the most optimistic estimate for the radius, we choose $\chi(X)$ negative and take its absolute value as small as possible, $\chi(X)=-2$. Moreover, we define the Calabi-Yau radius by the 6-sphere formula, ${\cal V} (2\pi\sqrt{\alpha^\prime})^6=(16/15)\pi^3R_\text{CY}^6$. Then the condition that the correction is large, ${\cal V}=\xi$, translates to $R_\text{CY}^6=15\zeta(3)\alpha^{\prime 3}/4$, in reasonable agreement with the previous estimate of how small a radius can be tolerated. We note, however, that the specific BBHL correction we used is not the one relevant for our actual case of interest, where a 3-sphere in a Calabi-Yau of large volume shrinks to small size. We are hence implicitly assuming that the BBHL correction provides a generic estimate of the typical string-frame radius at which $\alpha'$ corrections become important. The naive expectation $R\sim \sqrt{\alpha'}$, without any factors of $\pi$, appears to be confirmed.

Our small exercise demonstrates that, on the one hand, the naive estimates of the numerical prefactors do not lead to something as dangerous as a $(2\pi)^6$ on one side of the inequality. On the other hand, it is clear that for the concrete question whether all KKLT-type scenarios must fall victim to the singular-bulk or a similar problem, these prefactors may matter. One would need to study proper Calabi-Yau geometries to make progress.

\bibliographystyle{utphys}
\bibliography{groups}

\end{document}